\documentclass[prb,twocolumn,superscriptaddress,showpacs]{revtex4}
\usepackage{graphicx, color, epsf,bm,bbold}
\usepackage{grffile} 

\newcommand{\bea}{\begin{eqnarray}}
\newcommand{\eea}{\end{eqnarray}}
\newcommand{\beq}{\begin{equation}}
\newcommand{\eeq}{\end{equation}}
\newcommand{\vS}{\mathbf{S}}
\newcommand{\ud}{\mathrm{d}}
\newcommand{\Tr}{\mathrm{Tr}}
\newcommand{\pd}{\partial}
\newcommand{\cross}{\times}
\newcommand{\nenp}{Ni(C$_2$H$_8$N$_2$)$_2$NO$_2$ClO$_4\,$}

\begin{document}
\title{Frustration and Multicriticality in the Antiferromagnetic Spin-$1$ Chain} 
\author{J.\ H.\ Pixley\footnote{Present Address:
   Condensed Matter Theory Center, Department of Physics,
University of Maryland, College Park, Maryland 20742-4111, USA } }
\affiliation{Department of Physics and Astronomy, Rice University,
Houston, Texas, 77005, USA}
\author{Aditya Shashi}
\affiliation{Department of Physics and Astronomy, Rice University,
Houston, Texas, 77005, USA}
\affiliation{Department of Physics, Harvard University, Cambridge, Massachusetts 02138, USA}
\author{Andriy H. Nevidomskyy\footnote{nevidomskyy@rice.edu}}
\affiliation{Department of Physics and Astronomy, Rice University,
Houston, Texas, 77005, USA}

\date{\today}
\begin{abstract}
The antiferromagnetic spin-1 chain has a venerable history and has been thought to be well understood. Here we show that inclusion of both next nearest neighbor ($\alpha$) and biquadratic ($\beta$) interactions results in a rich phase diagram with a multicritical point that has not been observed before. We study the problem using a combination of the density matrix renormalization group (DMRG), an analytic variational matrix product state wavefunction, and conformal field theory. For negative $\beta < \beta^\ast$, we establish the existence of a spontaneously dimerized phase, separated from the Haldane phase by the critical line $\alpha_c(\beta)$ of second-order phase transitions. In the opposite regime, $\beta > \beta^\ast$, the transition from the Haldane phase becomes first-order into the next nearest neighbor (NNN) AKLT phase. Based on the field-theoretical arguments and DMRG calculations, we find that these two regimes are separated by a multicritical point $(\beta^\ast, \alpha^\ast)$ of a different universality class, described by the level-4 $SU(2)$ Wess--Zumino--Witten conformal theory. From the DMRG calculations we estimate this multicritical point to lie in the range $-0.2<\beta^\ast<-0.15$ and $0.47<\alpha^\ast < 0.53$. We further find that the dimerized and NNN-AKLT phases are separated from each other by a line of first-order phase transitions that terminates at the multicritical point. We establish that transitions out of Haldane phase into dimer or NNN-AKLT phase are topological in nature and occur either with or without closing of the bulk gap, respectively. 

We also study short-range incommensurate-to-commensurate transitions in the resulting phase diagram. Inside the Haldane phase, we show the existence of two incommensurate crossovers: the Lifshitz transition and the disorder transition of the first kind, marking incommensurate correlations in momentum and real space, respectively.  Notably, these crossover lines stretch across the entire $(\beta,\alpha)$ phase diagram, merging into a single incommensurate-to-commensurate transition line for negative  $\beta\lesssim \beta^\ast$ inside the dimer phase.  This behavior is qualitatively similar to that seen in classical frustrated two dimensional spin models, by way of the quantum $(1+1)D$ to classical $2D$ correspondence.\end{abstract}

\maketitle

\section{Introduction}

Quantum spin chains prominently display new phases and quantum phase transitions that emerge from the interplay of quantum and thermal fluctuations, geometrical frustration, and strong interactions.\cite{Diep.book} A variety of compounds can be modeled as quantum spin chains, such as \nenp (NENP) [Refs.~\onlinecite{Ma.1992,Hagiwara.1990}] and CsNiCl$_3$ [Ref.~\onlinecite{Kenzelmann.2002}] which realize antiferromagnetic spin $S=1$ Heisenberg chains, as well as $f$-electron compounds, like Yb$_4$As$_3$, representing the one dimensional spin-$1/2$ antiferromagnet~\cite{Kohgi.1997}. More recently in systems of ultra-cold atoms, there are now proposals to realize spin chains using spinor atoms in an optical lattice~\cite{Yip.2003,Imambekov.2003,Ripoli.2004}.  It is crucial to understand the role of geometrical frustration and competing interactions, which can be engineered artificially in cold atom systems, and appear naturally in the strong coupling approach to  the high temperature copper oxide and iron pnictide superconductors. For instance, it was shown that doping away from half-filling introduces frustration into the superexchange spin interaction~\cite{Inui.1988} in the cuprates, whereas the biquadratic spin-spin interaction $(\vS_i\cdot\vS_j)^2$  arises naturally  within the spin $S=1$ Heisenberg model describing the strong coupling theory of the iron pnictides~\cite{Yu.2012}.

We know from Haldane's mapping of the spin chain onto the non-linear sigma model~\cite{Haldane.1983} that the behavior of half-integer and integer spin-$S$ chains are dramatically different: the former possess gapless spinon excitations and quasi-long-range order~\cite{giamarchi.book} , whereas the spectrum of the latter is gapped with only short-range magnetic order. While the role of frustration has been studied extensively in the half-integer (in particular $S=1/2$) spin chains, its role in the integer spin case has been less well studied. Naturally, the simplest integer case is the $S=1$ spin chain, and this is the central focus of the present work, motivated in part by the aforementioned experimental realizations~\cite{Ma.1992, Kenzelmann.2002}. Another motivation of this work is to determine whether it is possible to realize a quantum phase transition between different phases. 
Previous studies of $S\!=\!1$ spin chains have revealed quantum phase transitions only at isolated points in the parameter space~\cite{Takhtajan.1982, Babudjian.1982,Uimin.1970,Kolezhuk.1996, Kolezhuk.1997-II}, which require fine tuning of the Hamiltonian and are therefore difficult to realize experimentally. In contrast to previous findings, as we will show in the present manuscript, the phase diagram of the frustrated spin-1 chain contains lines of phase transitions (both first and second order) that are accessible by moderate amount of magnetic frustration,  thus raising an exciting prospect of being observed experimentally.

\begin{figure}[t!]
\includegraphics[scale=.6]{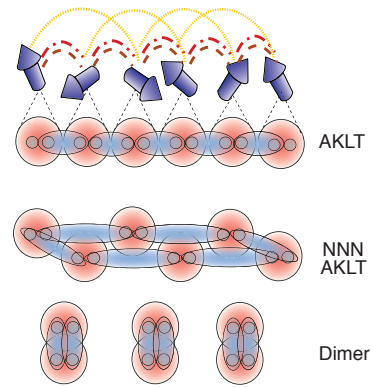}
\caption{(Color online) Schematic rendering of spin-$1$ chain with nearest (brown dashed), next-nearest (yellow dashed), and biquadratic (red dashed) interactions. Three ``typical'' ground states (GS) illustrated corresponding to the AKLT state, which is the exact GS at $\beta = 1/3,\alpha=0$, the ``next-nearest-neighbor'' AKLT (NNN-AKLT) which takes over for large $\alpha$, and the dimerized states characteristic of $\beta < -1$. We complement numerical studies of the model with an analytic variational wavefunction represented as a matrix product state which interpolates the three states described above.}
\label{fig:schem-states}
\end{figure}

Perhaps the simplest frustrated  model is the isotropic Heisenberg quantum spin chain with antiferromagnetic nearest neighbor and next nearest neighbor (NNN) interactions. Previous  DMRG~\cite{Schollwock.2005,Rizzi.2008} studies~\cite{Kolezhuk.1996, Kolezhuk.1997-II} of such a frustrated $S=1$ chain indicate that there is a first-order phase transition from the Haldane phase into the so-called next-nearest neighbor Affleck--Kennedy--Lieb--Tasaki (NNN-AKLT) phase characterized by singlet links along the NNN bonds, see Fig.~\ref{fig:schem-states}. Both the Haldane and NNN-AKLT phases are gapped, and the scenario of the first-order phase transition agrees with the field-theoretical analysis~\cite{Allen.1995} concluding that the spectral gap does not close for arbitrary values of the next-nearest neighbor interaction. 
 Another way to introduce frustration is by adding a competing biquadratic spin-spin interaction $(\vS_i\cdot\vS_j)^2$ into the model.  It turns out that this results in a significantly richer phase diagram harboring a variety of quantum phases, both gapped and gapless, which have been studied extensively in the past~\cite{Takhtajan.1982, Babudjian.1982,Uimin.1970,Affleck.1987,Nijs.1989, Kennedy.1992, Kennedy.1990, Barber.1989, Klumper.1989,Xian.1993,Reed.1994,Itoi.1997,Lauchli.2006,Schmitt.1998,Schollwock.1996,Bursill.1995,Sorenson.1990,AH.1987}. However, to the best of our knowledge, the combined effect of both the NNN and biquadratic interactions has not been studied previously,  and this is the subject of the present work. 
We therefore consider the most general spin $S=1$ chain with both types of frustration present:
\begin{equation}
H\!=\! J_1\left[\sum_i {\bf S}_i\cdot{\bf S}_{i+1}+\alpha\,{\bf S}_i\cdot{\bf S}_{i+2}+\beta({\bf S}_i\cdot{\bf S}_{i+1})^2\right]
\label{eqn:ham}
\end{equation}
We shall only consider the antiferromagnetic case (\mbox{$J_1\!>\!0$}, \mbox{$\alpha>0$}), with the aim to study the effect of frustration in the region $-1 \leq \beta<1$.  Below we shall first summarize the known theoretical results for the spin-$1$ chain in this regime.

The bilinear-biquadratic model with $\alpha=0$ has been studied extensively. As already mentioned, the isotropic NN Heisenberg spin model ($\beta=0$) lies in the \emph{Haldane phase}, characterized by exponentially decaying spin correlations and a gapped excitation spectrum, in stark contrast to antiferromagnets in higher dimension which display long range order and gapless spin-wave excitations, or half-integer spin chains as we have previously mentioned. 
The Haldane phase turns out to be stable for finite values of $\beta$ in the range $-1<\beta<1$, with a singlet ground state in the thermodynamic limit~\cite{Nijs.1989} and a finite string order parameter signifying the spontaneous breakdown of the hidden $Z_2\cross Z_2$ symmetry~\cite{Kennedy.1992}. 
In an open chain, the ground state is four-fold degenerate, due to the first excited triplet of the effective spin-1/2 edge excitations (``Kennedy triplet''\cite{Kennedy.1990}) becoming degenerate with the singlet ground state.
These edge excitations are the simplest example of gapless edge states, which 
are a consequence of the topologically non-trivial ground state in the bulk.  
A special case of $\beta=1/3$ has been analyzed by  Affleck, Kennedy, Lieb and Tasaki (AKLT)~\cite{Affleck.1987}, who found an exact ground state to be of the valence-bond-solid (VBS) type.  In the following, we refer to this as the AKLT point. It turns out that this point also marks the onset of incommensurability in the real-space spin-spin correlations, and falls into the classification of the \emph{disorder transition of the first kind}~\cite{Schollwock.1996} (see sections \ref{sec:sr-phases} and \ref{sec:disorder} for more details).   For larger values of $\beta$, one finds a Lifshitz transition at $\beta_L=0.43806(4)$ [Ref.~\onlinecite{Bursill.1995}] where the spin structure factor $S(q)$  develops a double-peak structure at incommensurate momenta. 

For  $\beta<-1$ and $\alpha=0$, the system is in the dimerized ground state (see  Fig.~\ref{fig:schem-states}), which is two-fold degenerate in the thermodynamic limit due to the two different possible dimer coverings of the chain. The dimer state has a finite gap~\cite{Barber.1989, Klumper.1989,Xian.1993,Sorenson.1990} (see Fig.~\ref{fig:schem-states}) and has a unique ground state for an open chain with even number of sites. 
 The point $\beta=-1$ at which the transition occurs turns out to be exactly solvable~\cite{Takhtajan.1982, Babudjian.1982} and is known as the Takhtajan--Babudjian (TB) point, described by the critical SU$(2)_{k=2}$  Wess--Zumino--Witten (WZW) conformal field theory~\cite{AH.1987}.
On the positive $\beta$ side, the Haldane phase is flanked by a gapless phase~\cite{Reed.1994,Itoi.1997} with antiferro-quadrupolar quasi-long range order~\cite{Lauchli.2006} for $\beta>1$.  The transition between the two is marked by the exactly solvable Uimin--Lai--Sutherland (ULS) point~\cite{Uimin.1970} at $\beta=1$.  In this work, we limit the discussion to the dimerized and Haldane phases and do not consider the regime $\beta>1$. The effect of frustration on the ULS point due to NNN interactions is left for future study.

As the reader can see, the spin-1 chain with $\alpha=0$ has a venerable history. The effect of NNN interactions ($\alpha>0$) on the other hand, has scarcely been studied. To the best of our knowledge, only the case of vanishing biquadratic interaction $\beta=0$ has been addressed. In this case, the  authors of Refs.~\onlinecite{Kolezhuk.1996,Kolezhuk.1997-II} found, using a combination of DMRG calculations and an analytic variational wavefunction approach, that the Haldane phase is stable for $\alpha$ in the range $\alpha < \alpha_T$ and that  the AKLT valence-bond state wavefunction still provides a good description of the ground state. We will therefore refer to this VBS state as the ``AKLT state'' in what follows, even though it spans a much broader region around the AKLT point $(\beta=\frac{1}{3}, \alpha=0)$ originally studied in Ref.~\onlinecite{Affleck.1987}.
Similar to the pure biquadratic model ($\alpha=0$),  disorder and Lifshitz points were found in the NNN Heisenberg model ($\beta=0$) at $\alpha_d=0.284(1)$ and $\alpha_L=0.3725(25)$, respectively~\cite{Kolezhuk.1996,Kolezhuk.1997-II} (see section~\ref{sec:disorder} for more detail).  Intriguingly, for larger values of $\alpha$ along the $\beta=0$ axis, there is a phase transition at $\alpha_T=0.7444(6)$ from the AKLT phase to the so-called next nearest neighbor AKLT (NNN-AKLT) phase~\cite{Kolezhuk.1996,Kolezhuk.1997-II} illustrated in Fig.~\ref{fig:schem-states}.  At the transition, the DMRG calculations showed~\cite{Kolezhuk.1996,Kolezhuk.1997-II} that the gap remains finite, in agreement with earlier field-theoretical calculations.~\cite{Allen.1995} Although the DMRG calculations in Refs.~\onlinecite{Kolezhuk.1996,Kolezhuk.1997-II} do not find a discontinuous first derivative of the ground state energy at the transition, the  authors  concluded the transition to be first order, based on the finite jump of the string order parameter at $\alpha_T$, while the bulk gap remains open across the transition. In the following, whenever we refer to a first order transition between the AKLT and NNN-AKLT phases, this is the transition we have in mind.   The disappearance of the  string order parameter above $\alpha_T$, together with the gapping out of the edge excitations, imply that this transition is topological in nature, from the topological AKLT phase into a topologically trivial NNN-AKLT phase where the $Z_2\cross Z_2$ symmetry is restored~\cite{Kolezhuk.1996, Kolezhuk.1997-II}. The topological aspect of the transition is discussed in more detail in Section~\ref{sec:edge}.

The above findings along the $\beta=0$ axis suggest that the NNN interaction $\alpha$ has a profound effect on the ground state of the spin-1 chain, yet virtually nothing is known about its phase diagram when both $\alpha$ and  $\beta$ are finite. In this work, we aim to fill in this gap while attempting to address the following main questions:
\begin{itemize}
\item What is the topology of the phase diagram in the $(\beta,\alpha)$ plane? Are all transitions first order as in the $\beta=0$ case? \vspace{-1.5mm}
\item  What happens in the vicinity of the gapless TB point ($\beta\!=\!-1, \alpha\!=\!0$) for finite $\alpha$? Does the gapless behavior survive for a range of $\alpha$, or is an  infinitesimal value of $\alpha$ sufficient to gap out the spectrum? 
\vspace{-3mm}
\item How are the dimerized and the NNN-AKLT phases connected? Is there a phase transition between the two? \vspace{-1.5mm}
\item How does the incommensurability of spin-spin correlations set in? What is the significance of the AKLT point ($\beta=1/3$) in the presence of a finite NNN interaction $\alpha$?  \vspace{-1.5mm}
\item Can one obtain a realistic description of the system using an analytic variational wavefunction based on matrix product states by keeping only four states, shown to work successfully~\cite{Kolezhuk.1997-II} for $\beta=0$?
\end{itemize}

The remainder of the paper is organized as follows.  We first state our main results as revealed in the calculated phase diagram and briefly describe various phases and transitions between them in section~\ref{sec:phase-diagram}.  
Next, we discuss the emergence of a line of critical points in the vicinity of the TB point, whose existence we verify using conformal field-theory arguments, in section~\ref{sec:critical-line} . We then discuss each phase comprehensively, and outline the details of the DMRG and variational wavefunction approaches used to obtain them in section~\ref{sec:phases}. The nature of the incommensurability in the spin-spin correlations and resulting short-range phases are examined in detail in section~\ref{sec:disorder}.  We then discuss the results and future directions in section~\ref{sec:discussion}. We conclude with a summary of the results in section~\ref{sec:conclusions} and section~\ref{sec:methods} contains the technical details of the various methods used in this work.

\section{Phase Diagram}
\label{sec:phase-diagram}

\subsection{Thermodynamic phases and transitions}
\begin{figure}[!t]
\includegraphics[width=3.in]{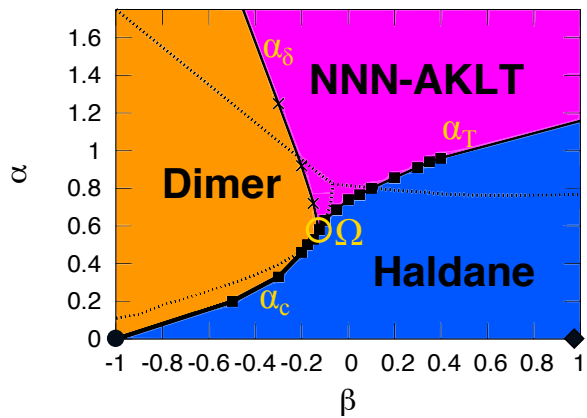}
\caption{(Color online) Schematic phase diagram obtained in this work. Symbols marking the transitions between different phases are the DMRG data, solid lines are guide to the eye, and dashed lines are obtained from the analytical matrix-product state ansatz (see section \ref{sec:MPS}). The critical line (thick black line) $\alpha_c(\beta)$ starts at the TB point $\beta=-1$ (solid circle), and is terminated at a multi-critical point $\Omega$, beyond which the transition becomes first order (thin black line $\alpha_T$).  The critical line $\alpha_c$ separates the dimer and Haldane phases whereas first-order transition $\alpha_T$ separates the Haldane and NNN-AKLT phases. The transition between the dimer and NNN-AKLT phases is of first order, marked by $\alpha_\delta$ line (crosses). We have not studied the effect of $\alpha$ on the ULS point (solid diamond) and the region $\beta \geq 1$. }
\label{fig:schem-pd}
\end{figure}

Our results can be most clearly stated by first presenting a schematic phase diagram in Figure~\ref{fig:schem-pd}.
 Using a combination of field theoretical arguments and DMRG calculations guided by an analytic MPS wavefunction approach, we find the existence of a critical line of the 
 $2^{\text{nd}}$ order transitions  between the dimer and Haldane phases, denoted as         $\alpha_c(\beta)$ in Figure~\ref{fig:schem-pd}.  The critical line starts at the TB point and terminates at a multicritcal point $\Omega=(\beta^*,\alpha^*)$ turning into a line of first order transitions $\alpha_T(\beta)$,  passing through the point $[\beta=0,\alpha_T=0.7444(6)]$ which was previously studied with DMRG~\cite{Kolezhuk.1996,Kolezhuk.1997-II}.  We find that the critical line $\alpha_c(\beta)$ separates the Haldane phase from the dimer phase which breaks the translational symmetry of the lattice.
 The first-order transition line $\alpha_T(\beta)$, on the other hand, separates the Haldane phase below from the NNN-AKLT phase above $\alpha>\alpha_T$, where the broken $Z_2\cross Z_2$ symmetry is restored. This line of first-order transitions extends into the region $\beta^*< \beta<1$, and in the following we refer to this line simply as the ``transition'' line $\alpha_T$ (as opposed to the``critical'' line $\alpha_c$). 
 The NNN-AKLT phase does not break translational symmetry of the lattice (for more detail, see section~\ref{sec:discussion}), and as a result is distinct from the dimer phase with a first order phase transition separating the two along the $\alpha_{\delta}$ line. Lastly, both the dimer and NNN-AKLT phases are distinct from the Haldane phase in the topological sense: they are both topologically trivial and unlike the Haldane phase, do not possess zero-energy edge excitations (see section~\ref{sec:edge}).

In addition to DMRG, we have used an analytical matrix product state (MPS) ansatz (see Section~\ref{sec:MPS} for the details of the method) to gain a semi-quantitative understanding of various phases and transitions between them.  We followed the authors of Ref.~\onlinecite{Kolezhuk.1997-II} in proposing the analytic variational MPS ansatz for the trial wavefunction that keeps $M=4$ states and interpolates between the AKLT, NNN-AKLT, and dimer ground states. 
The relative stability of these phases across the $(\alpha,\beta)$ phase diagram is judged based on the lowest variational MPS ground state energy. As a result, all transition lines in this approach (thin dashed black lines in Fig.~\ref{fig:schem-pd}) are first order, corresponding to the level crossing of respective ground state energies. Despite its shortcomings, the phase diagram agrees qualitatively with the phase diagram calculated by DMRG (see the dashed lines in Fig.~\ref{fig:schem-pd}). Moreover, in the approach to $\beta=0$ from the left, the agreement is even quantitatively reasonable. Note however, that the variational approach underestimates the value of $\alpha_T(\beta)$  for $\beta>0$, and predicts an incorrect decreasing trend for $\alpha_T(\beta)$ for positive $\beta$ (compare with the DMRG transition line in Fig.~\ref{fig:schem-pd}). 

It is important to note that the DMRG can also be formulated in the language of MPS which it variationally optimizes~\cite{Schollwock.2011, Versraete.2008} while keeping a very large number of states $M$ that cannot be handled analytically. Therefore the DMRG is far more accurate than the simple analytic MPS ansatz with $M=4$ which we employed, however the  latter is still a powerful tool as it allows us to gain valuable qualitative insights into the transitions between different phases.  

\subsection{Short-range order} \label{sec:sr-phases}
\begin{figure}[t!]
\includegraphics[width=3.25in]{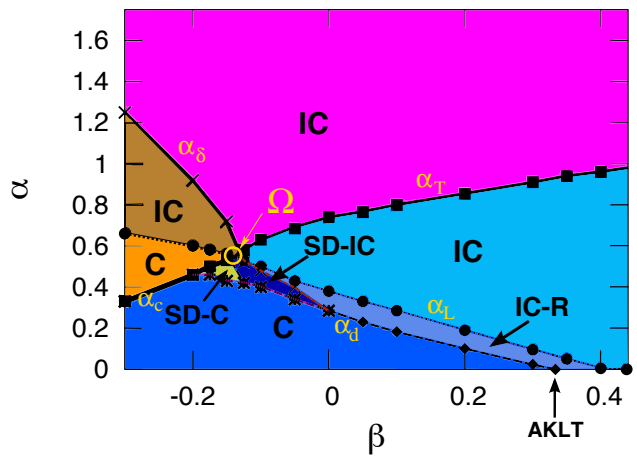}
\caption{(Color online) DMRG phase diagram indicating short-range order within the phases illustrated earlier in Fig.~\ref{fig:schem-pd}. Various commensurate (C)  and incommensurate (IC) phases are described in the text. 
Note that the onset of real-space incommensuration (IC-R) across the $\alpha_d$ line (diamonds) is distinct from the Lifshitz line $\alpha_L$ (circles) inside the Haldane phase. The two lines merge into a single C-IC transition upon entering the long-range dimer  phase at $\beta \lesssim -0.15$. 
}
\label{fig:quant-pd}
\end{figure}

In addition to the aforementioned three distinct phases (Haldane, NNN-AKLT and dimer), we also used the DMRG to identify regions of the phase diagram characterized by various types of short-range order through the real-space spin-spin correlation functions  $C(x) = \langle {\bf S}(x)\cdot {\bf S}(0)\rangle$. These short-range ordered phases are not true thermodynamic phases as they lack long-range order, however they featured promptly in earlier studies by other authors and are important for understanding the evolution of the correlations between different regions of the phase diagram. 
Guided by the well-known mapping of the quantum spin chain to the two dimensional classical spin model, the correlation function $C(x)$ is well described by the two-dimensional ($d=2$) Ornstein--Zernicke (OZ) form:
\begin{equation}
C_{OZ}(x) \propto \displaystyle \cos[q(\alpha,\beta)\cdot x]\;\frac{e^{-x/\xi(\alpha,\beta)}}
{x^{(d-1)/2}},
\label{eqn:OZ}
\end{equation}
adopted in the previous studies of the spin-1 Heisenberg chain~\cite{White.1993,Schollwock.1996,Kolezhuk.1996,Kolezhuk.1997-II}. The Fourier-transform of $C(x)$ defines the momentum-space correlation function
\begin{equation}
S(q) = \sum_{x}e^{iqx}C(x).  
\label{eqn:S(q)}
\end{equation}
Various short-range commensurate (C) and incommensurate (IC) phases are shown in Fig.~\ref{fig:quant-pd}, while the detailed DMRG analysis of their properties is deferred to Section~\ref{sec:disorder}: 
\begin{description}
\item[Haldane phase]
\hfill \\
$\bullet$ {\bf C:}  Short range antiferromagnetic correlations, with a real space correlation function, $C(x)$ well described by the OZ form in Eq.~(\ref{eqn:OZ}) with $q(\alpha,\beta)=\pi$.  The momentum space correlation function, $S(q)$ has a single peak at $\pi$.
\newline
$\bullet$ {\bf IC-R:} Short range antiferromagnetic correlations with incommensurate correlations in real space.  $C(x)$ is well described by the OZ form with $q(\alpha,\beta)>\pi$ and $S(q)$ has a single peak at $\pi$.  In this phase, the ground state is closest to the AKLT wavefunction ansatz.
\newline
$\bullet$ {\bf IC:}  Short range antiferromagnetic correlations with incommensurate correlations in both, real and momentum space.  $C(x)$ is well described by the OZ form with $q(\alpha,\beta)>\pi$.  S(q) has two symmetric peaks at an incommensurate wave vector, see Figure~\ref{fig:chiq}.
\newline
$\bullet$ {\bf SD-C:}  Short range dimer phase with commensurate spin correlations. 
While the dimer order parameter is zero (no long-range order), 
a short range dimer order is manifest in the spin-spin correlation function by the appearance of a finite dimerization $\delta(\alpha,\beta)>0$ on top of the OZ form Eq.~(\ref{eqn:OZ}):
\begin{equation}
C_D(x) \propto \displaystyle (1+\delta(\alpha,\beta)(-1)^x)C_{OZ}(x).
\label{eqn:OZD}
\end{equation}
The spin correlations are commensurate, with $q(\alpha,\beta)=\pi$.  The momentum space correlation function, $S(q)$ also has a single peak at $\pi$. 
\newline
$\bullet$ {\bf SD-IC:} Short range dimer phase, similar to SD-C, except the real-space spin-spin correlations become incommensurate, as characterized by $q(\alpha,\beta)\neq\pi$ in Eqs.~(\ref{eqn:OZ}) and (\ref{eqn:OZD}).
\item[Dimer]
\hfill \\
$\bullet$ {\bf C:} Translational symmetry broken, the dimer order parameter is finite in the thermodynamic limit, $C(x)$ is reasonably well described by the dimerized OZ form, Eq.~(\ref{eqn:OZD}) with $\delta(\alpha,\beta)>0$ and $q(\alpha,\beta)=\pi$.  The ground state energy is very close to the dimer wavefunction ansatz and $S(q)$ has a single peak at $\pi$.
\newline
$\bullet$ {\bf IC:} Translational symmetry broken, with short range incommensurate correlations, the dimer order parameter is finite in the thermodynamic limit, $C(x)$ is reasonably fitted to the dimerized OZ form, Eq.~(\ref{eqn:OZD}) with $\delta(\alpha,\beta)>0$ and $q(\alpha,\beta)>\pi$.  $S(q)$ has two symmetric peaks at an incommensurate wave vector.
\item[NNN-AKLT]
\hfill \\
 Translational symmetry is restored, with a negligible dimer order parameter in the thermodynamic limit. The entire phase is incommensurate since spin-spin correlation function $C(x)$ is well described by the OZ form with $q(\alpha,\beta)>\pi$.  $S(q)$ has two symmetric peaks at an incommensurate wave vector $q$ that approaches $\pi/2$ for large values of $\alpha$.
\end{description}  

\noindent
Several remarks are in order. First, inside the short-range ordered Haldane phase, the incommensuration in the real-space correlation function Eq.~(\ref{eqn:OZ}) occurs via the so-called disorder transition of the first kind well documented in classical statistical mechanics\cite{Stephenson} and first noted in the $S=1$ chain by Schollw\"ock \emph{et al}.\cite{Schollwock.1996} Previous studies have focused on isolated points along either $\alpha=0$ line\cite{Schollwock.1996}, where the disorder transition coincides with the AKLT point at $\beta=1/3$, or along the $\beta=0$ line, where the disorder transition was found\cite{Kolezhuk.1996,Kolezhuk.1997-II} at $\alpha_d=0.284(1)$. In this work, we have mapped out the entire phase diagram in the $(\alpha,\beta)$ plane and found that these two points are connected in a smooth fashion by a line of the disorder transitions $\alpha_d(\beta)$ which separates the commensurate (C) phase below the line from the IC-R phase above it, see Fig.~\ref{fig:quant-pd}.
 Interestingly, a phenomenon of dimensional reduction occurs at the disorder transition~\cite{Schollwock.1996}, whereby the system becomes effectively one-dimensional $d\!=\!1$ (rather than generically $1\!+\!1$ dimensional, $d\!=\!2$). This is manifest in the change of the power-law dependence  in the denominator of the OZ equation (\ref{eqn:OZ}) from $x^{1/2}$ to becoming $x$-independent along the disorder line $\alpha_d(\beta)$.

Second, in the Haldane phase it turns out that this disorder transition is distinct from the Lifshitz transition  where the momentum-space structure factor $S(q)$ acquires an incommensurate peak at $q\neq \pi$, as first documented in Ref.~\onlinecite{Schollwock.1996}. Upon increasing $\alpha$ or $\beta$, the  commensurate (C) phase first undergoes the disorder transition into the IC-R phase, before crossing the Lifshitz line $\alpha_L(\beta) > \alpha_d(\beta)$ into the IC phase, as shown in Fig.~\ref{fig:quant-pd}. We find that these two distinct lines merge into a single commensurate-to-incommensurate transition upon entering the translational-symmetry broken dimer phase, similar to what has been seen in frustrated classical two dimensional Heisenberg models~\cite{Garel.1986}. For more details, see Fig.~\ref{fig:xiandq}(c) in section~\ref{sec:disorder} and discussion therein.
  Intriguingly, to the accuracy of our finite-size DMRG calculations, we find the location of the multi-critical point $\Omega$ to lie at the intersection of the disorder $\alpha_d$, Lifshitz $\alpha_L$, and critical $\alpha_c$ lines, as Figure~\ref{fig:quant-pd}  illustrates.

\section{Critical Line Between Haldane and Dimer phase}
\label{sec:critical-line}
\subsection{Field Theoretical Picture} 
\label{sec:field-theory}
\subsubsection{Non-Abelian bosonization of S=1 chain}
We first focus on the phase boundary between the dimer and Haldane phases, shown as the thick solid black line in Fig.~\ref{fig:schem-pd}. 
It is instructive to consider the vicinity of the integrable TB point ($\beta=-1$), which is equivalent in the continuum limit to a level-2 SU(2) Wess-Zumino-Witten (WZW) model~\cite{AH.1987}. The fundamental question is what happens to this critical point in the phase space of parameters $(\beta,\alpha)$ of the Hamiltonian Eq.~(\ref{eqn:ham}). It would seem that there could be two possibilities: (a) either both the next-nearest neighbour interaction $\alpha (\vS_i\cdot \vS_{i+2})$ and the biquadratic interaction  $\beta (\vS_i\cdot \vS_{i+1})^2$ turn out to be relevant and thus open up a gap in the spectrum of excitations, or (b) there is an irrelevant direction in the ($\beta,\alpha$) plane that leaves the system critical, in which case the resulting theory ought to be described by 
a conformal field theory. Which of these two possibilities is realized?

 In order to answer this question, we must understand how the integrable TB point can be perturbed, and for this we must consider all symmetry-allowed perturbations of the corresponding WZW theory. This seemingly impossible task is achievable by virtue of the conformal field theory, which dictates that all perturbing fields can be expressed in terms of the primary conformal fields. 
The SU(2)$_2$ theory is characterized by two non-trivial representations of bilinears built from primary fields: the doublet $g_{mn}$ with conformal dimensions $\left(\frac{3}{16},\frac{3}{16}\right)$, which transforms in the fundamental representation of SU(2), and the triplet $\Phi_{ab}$ with conformal dimensions  $\left(\frac{1}{2},\frac{1}{2}\right)$, which transforms in the 3-dimensional adjoint representation. 
 All possible perturbations to the WZW theory can be constructed from these two scaling fields and their conformal descendants, in addition to the bilinear of primary currents $(J^a \bar{J}^a)$. It was shown by Affleck and Haldane ~\cite{AH.1987} that the scaling field $g_{mn}$ is proportional to the staggered magnetization $(-1)^i \vS_i\cdot \vS_{i+1}$ and is therefore not allowed to appear as a perturbation, because it would otherwise explicitly break the translational symmetry of the chain. We thus conclude that small deviations from the TB point can be described, in the continuum limit, by the  Lagrangian of the SU(2)$_2$ WZW theory plus the following perturbations:
\beq
\mathcal{L} = \mathcal{L}_{\text{WZW}} + m \Tr (\hat{\Phi}) -\lambda J^a \bar{J}^a,
\label{eqn:lagrange}
\eeq
(summation implied over field components $a=1,2,3$).

The $\Tr(\hat{\Phi})$ term has scaling dimension $\Delta=1$ and is strongly relevant in (1+1) dimension, leading to the opening of the spin gap. Depending on the sign of the ``mass'' $m$, the system ends up either in the Haldane phase ($m<0$) or in the symmetry-breaking dimerized phase ($m>0$). In fact, for $\alpha=0$, the mass is proportional to the distance from the TB point~\cite{AH.1987}, $m \propto -(1+\beta)$. 

\begin{figure}[!tb]
\includegraphics[width=0.5\textwidth]{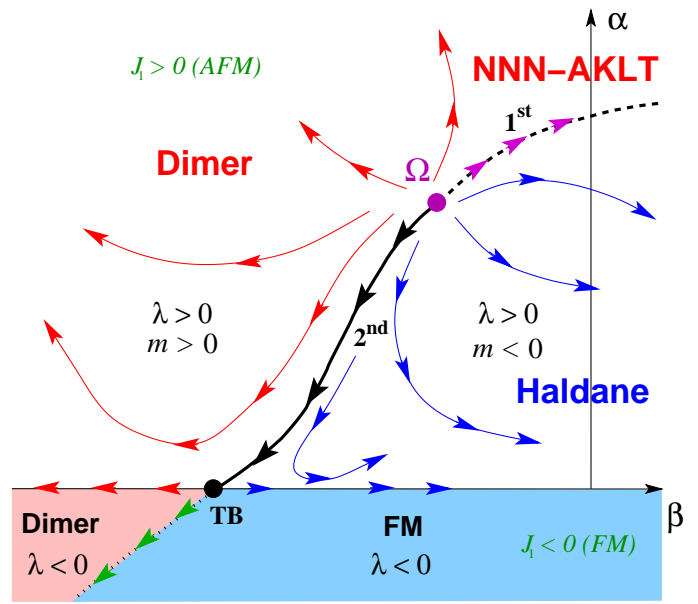}
\caption{(Color online) Schematic RG flow diagram in the vicinity of the Takhtajan--Babudjian (TB) point. The signs of conformal theory parameters $m$ and $\lambda$ in Eq.~(\ref{eqn:lagrange}) determine the nature of the phases (Dimer, Haldane, or ferromagnet). The blue and red arrows indicate the flow of the relevant parameter $m$ which opens up a spectral gap upon entering the Haldane or dimer phase, respectively. 
 The solid black line is critical,  and is governed by the conformal WZW theory at the TB point. The black arrows indicate the marginally irrelevant flow of $\lambda$ towards the TB point (if $J_1>0$ is antiferromagnetic), whereas the green arrows indicate marginally relevant flow (if $J_1<0$, not discussed in this work). We identify a new mutlicritical point $\Omega$, characterized by an unstable RG flow (purple arrows), marking the $1^{\rm st}$ order phase transition everywhere along the dashed line (see text for discussion).}
\label{fig:flow}
\end{figure}

Let us now consider the last term in Eq.~(\ref{eqn:lagrange}). It has a scaling dimension $\Delta=2$ and is marginal at tree level. Higher order RG calculations show~\cite{Cardy.1996} that this term is marginally irrelevant for $\lambda >0$ (corresponding to the antiferromagnetic sign of the nearest-neighbour interaction $J_1$), whereas for negative $\lambda$ (ferromagnetic $J_1$), it becomes marginally relevant, flowing in strong coupling towards either the gapped dimerized state or gapless ferromagnetic state. In this work, we shall only consider the case of antiferromagnetic $J_1$, so for our purposes $\lambda>0$ is always marginally irrelevant.

\begin{figure}[tb!]
\includegraphics[width=3.in]{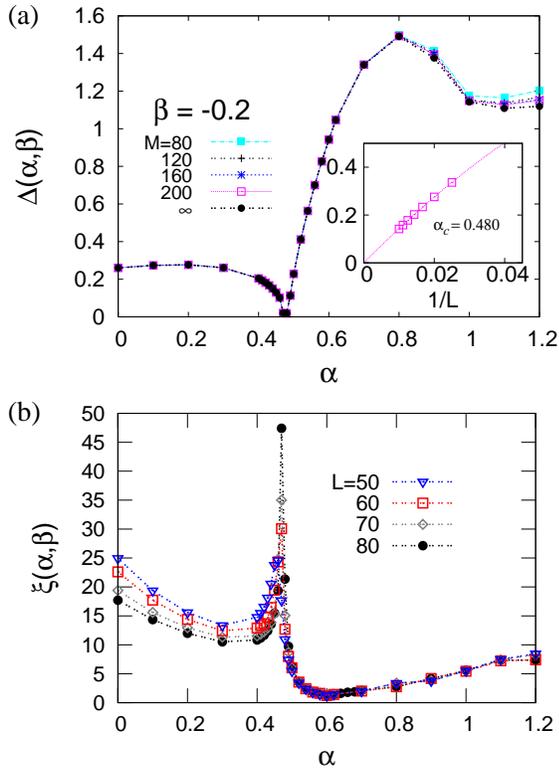}
\caption{(a) The bulk gap $\Delta(\alpha,\beta=-0.2)$ in the thermodynamic limit for  different number $M$ of kept DMRG states  as a function of $\alpha$.  Inset: At the critical point $\alpha_c=0.480(5)$,  we find a small gap $\Delta = 0.018(3)$ for the largest $L$ and $M$ values studied. However extrapolating to the infinite system size, we find an excellent fit to a gapless form $\Delta(L)=a_{\Delta}/L + b_{\Delta}/L^2$ for a fixed number of states $M=200$.  (b) The Correlation length for various system sizes $L$ with $M=200$ [see equation (\ref{eqn:OZ}) for a definition of $\xi$], as we move through the critical line, with a disorder point at the minimum of $\xi$ at $\alpha_d(\beta=-0.2)=0.600(5)$.}
\label{fig:xi-gap2}
\end{figure}

The values of the parameters $(m,\lambda)$ in the effective continuum theory (\ref{eqn:lagrange}) are related in some fashion to the parameters $(\alpha,\beta)$ of the original Hamiltonian, however the exact correspondence is, unfortunately, unknown. 
Nevertheless, the above scaling argument allows us to sketch the RG flow in the vicinity of the TB point, see Fig.~\ref{fig:flow}. The solid black line is massless ($m=0$) and remains critical for $\lambda>0$ at least for small perturbations, with the critical exponents governed by the flow toward the integrable WZW SU(2)$_2$ theory at the TB point (plus logarithmic corrections due to marginally irrelevant operators). Crossing the solid black line in Fig.~\ref{fig:flow} signifies the $2^{\text{nd}}$ order phase transition between the Haldane phase and the dimerized phase. This is indeed what our DMRG calculations show for $\beta\lesssim-0.2$, indicating a vanishing gap to the bulk excitations in the thermodynamic limit and diverging correlation length, see Fig.~\ref{fig:xi-gap2}. 
The information about the correlation length $\xi$ in Fig.~\ref{fig:xi-gap2}b) has been extracted from the OZ form of the real-space correlation function, Eq.~(\ref{eqn:OZ}).

\begin{figure}[tb!]
\includegraphics[width=3.in]{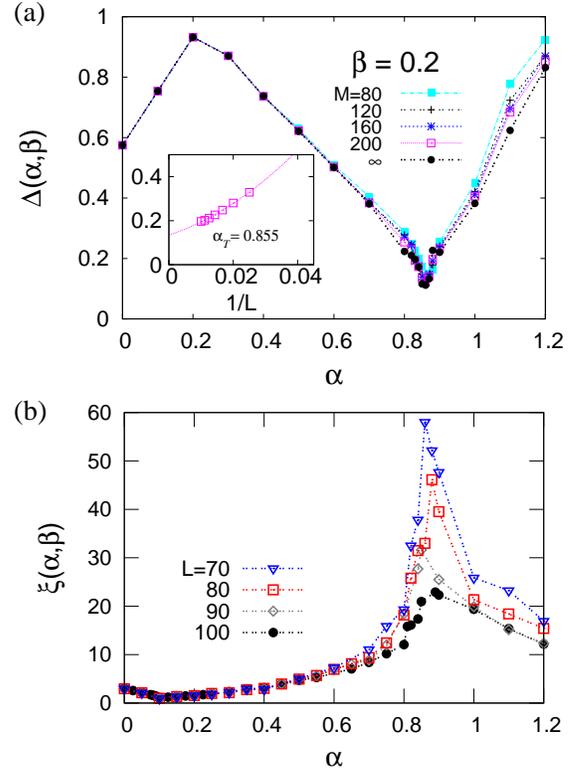}
\caption{(a) The bulk gap $\Delta(\alpha,\beta=0.2)$ in the thermodynamic limit for  different number $M$ of kept DMRG states as a function of $\alpha$. We find that the gap goes through a minimum in the vicinity of $\alpha_T=0.855(5)$, with a minimum value of $\Delta_0=0.11(2)$. Inset: At the transition point, we find an excellent fit to a gapped form $\Delta(L)=\Delta_0+a_{\Delta}/L + b_{\Delta}/L^2$ for a fixed number of states $M=200$.  (b) Correlation length for various system sizes $L$ with $M=200$ [see equation (\ref{eqn:OZ}) for a definition of $\xi$], as we move through the critical line and a disorder point at $\alpha_d(\beta=0.2)=0.100(5)$.}
\label{fig:xi-gap1}
\end{figure}

Intriguingly, in the absence of a biquadratic coupling, $\beta=0$, 
Kolezhuk and co-workers argued that upon increasing $\alpha$, the transition from the Haldane phase to the dimerized NNN-AKLT phase is first-order, with the bulk gap never closing across the transition~\cite{Kolezhuk.1996, Kolezhuk.1997-II}.  Our DMRG results corroborate this statement for $\beta\gtrsim -0.2 $, see Fig.~\ref{fig:xi-gap1}. How do we reconcile the seemingly contradictory results in Figs.~\ref{fig:xi-gap2} and \ref{fig:xi-gap1}? Could the transition be first order throughout the phase diagram with a small gap which is below the DMRG detection threshold? Conversely, could it be that the true bulk gap does vanish across the transition, however the finite-size effects result in a numerically finite gap, even when extrapolated to $L\to \infty$? Indeed, it is very difficult within DMRG to distinguish a system with a small but finite gap from a true gapless case, as the recent state-of-the-art DMRG studies on the kagom\'e lattice demonstrate \cite{Yan.2011, Depenbrock.2012}. Below we shall settle this ambiguity with the help of field-theoretical analysis. What we find is that the second-order transition for $\beta\lesssim -0.2$ (see Fig.~\ref{fig:xi-gap2}) and the apparent first-order transition for positive $\beta$ (Fig.~\ref{fig:xi-gap1}) are both correct, and that there exist a critical end-point $\Omega$ terminating the line of the second-order transitions in the phase diagram Fig.~\ref{fig:flow}.

\subsubsection{Multicritical point: Conformal field theory}\label{sec:CFT1}

Based on the DMRG results for the spectral gap and spin-spin correlation functions, we propose the existence of a critical end-point $\Omega$ in the phase diagram Fig.~\ref{fig:flow}. The continuum limit of this critical point must be described by a conformal field theory, however it becomes quickly apparent that this cannot be the same SU(2)$_2$ WZW theory that describes the TB point at $\beta=-1$. The fundamental reason is that as explained above, the  SU(2)$_2$ field theory has only one relevant operator: the mass term $m\Tr(\hat{\Phi})$ in Eq.~(\ref{eqn:lagrange}). By contrast, the critical end-point $\Omega$ must have two relevant (or marginally relevant) operators -- one playing the role of the mass term that opens up the Haldane gap (blue and red flow lines in Fig.~\ref{fig:flow}), and another one, which makes the critial point $\Omega$  unstable towards the 1$^{\rm st}$ order phase transition along the purple flow line in Fig.~\ref{fig:flow}. In other words,  $\Omega$ must be an unstable fixed point in the ($\alpha, \beta$) phase-space, and this requires more than one relevant operator. 

The sought conformal field theory must therefore be richer than the SU(2)$_2$ theory and must satisfy the following requirements: (a) it must possess at least two relevant (or marginally relevant) scaling fields, as explained above;  (b) it must satisfy Zamolodchikov's ``$c$-theorem''~\cite{Zamolodchikov.1986}, so that its central charge must monotonically \emph{decrease} to that of the TB point under the RG flow (along the thick black flow line in Fig.~\ref{fig:flow}). This latter constraint means that the central charge must be larger than $c_0=3/2$ of the SU(2)$_2$ theory.

\begin{figure}[t!]
\includegraphics[width=3.in]{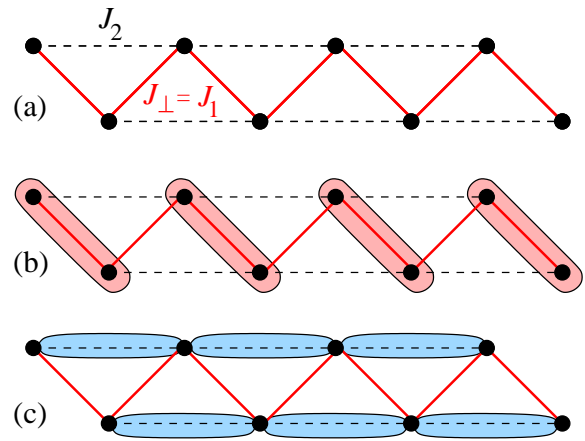}
\caption{(Color online) Schematic representation of two coupled $S=1$ spin chains. (a) The spin coupling within each chain is given by the interaction $J_2 = \alpha J_1$ in the original model Eq.~(\ref{eqn:ham}). (b) The dimerized ground state, with the circled rungs corresponding to the spin singlet (quintuplet) of two spins, for antiferromagnetic (ferromagnetic) interchain coupling $J_\perp = J_1$. (c) The NNN-AKLT ground state, with the blue links representing singlet bonds between composite spin-$1/2$ objects that form the AKLT ground state between second neighbors.}
\label{fig:chains}
\end{figure}

We show in the Section \ref{sec:CFT} that the simplest conformal field theory that satisfies these requirements is the SU(2)$_{k=4}$ WZW theory. The  SU(2) Lie group is natural because the Hamiltonian is generically SU(2)-symmetric (higher ``accidental'' symmetry is possible, but would require fine-tuning the parameters of the Hamiltonian). As for the level $k=4$, we give a formal argument based on conformal embedding in Sect.~\ref{sec:CFT}, but this can be understood qualitatively as follows. The Hamiltonian in Eq.~(\ref{eqn:ham}) can be represented as two $S=1$ spin chains, each with half as many sites, interacting via the intra-chain coupling $J_2 = \alpha J_1$, see Fig.~\ref{fig:chains}. The two chains are coupled by the Heisenberg as well as biquadratic spin-spin interaction, so that the Hamiltonian  $H=H_0 + H_\perp$:
\bea
H_0 &=& J_2 \sum_{\alpha=1}^2 \sum_i \vS_{\alpha,i}\cdot \vS_{\alpha,i+1} \label{eqn:chains}\\ 
H_\perp &=& J_\perp\sum_i\sum_{\delta=\pm1/2}   \vS_{1,i+1}\cdot\vS_{2,i+\delta} + \beta \left(\vS_{1,i+1}\cdot\vS_{2,i+\delta} \right)^2 \nonumber,
\eea
where $J_\perp = J_1$ in the original model Eq.~(\ref{eqn:ham}).
 In the limit $J_2 \gg J_1$ ($\alpha\gg 1$), the two chains are decoupled and each can be described by the perturbed WZW theory in Eq.~(\ref{eqn:lagrange}). However for finite $J_1$, the local staggered magnetizations of the two chains can lock together, forming a combined spin $S=2$ object. This is particularly evident in the case of a ferromagnetic coupling $J_1$, when the two chains form a ladder of quintuplets, leading to a dimerized ground state shown in Fig.~\ref{fig:chains}b. For antiferromagnetic intechain coupling $J_1$, the situation is similar but it is the singlets that form instead on the dimerized bonds. The bottom line is that emergent $S=2$ excitations can form. Provided these excitations are gapless at a (multi)critical point, it was  proposed by Affleck~\cite{Affleck.PRL.1986} that they  are described by SU(2)$_k$ WZW conformal field theory at level $k=2S$. This has been corroborated recently by numerical calculations~\cite{Thomale.2012}. 
In our case $S=2$ and the resulting theory is thus SU(2)$_4$. 

The  SU(2)$_4$ WZW theory has central charge $c=2$ and the RG flow towards the TB point (Fig.~\ref{fig:flow}) indeed satisfies Zamolodchikov's ``$c$-theorem''~\cite{Zamolodchikov.1986} mentioned earlier. The analysis shows that the theory has 4 primary fields, however only two of them satisfy the symmetries of the Hamiltonian and are therefore allowed as perturbations~\cite{ZF.1986, Gepner.1986}. These two primary fields have conformal dimensions ($\frac{1}{3},\frac{1}{3}$) and (1,1). The corresponding field bilinears have scaling dimensions  $\Delta_1=2/3$ (relevant) and $\Delta_2=2$ (marginally relevant). In addition, the current bilinear $\bar{J}^a\!J^a$ (scaling dimension $\Delta_3=2$) is also allowed, however it is marginally irrelevant for antiferromagnetic $J_1>0$ and leads only to logarithmic corrections to scaling. As per our requirement, the SU(2)$_4$   theory thus possesses two relevant fields, making $\Omega$ an unstable fixed point. The field with dimension $\Delta_1=2/3$ plays a role similar to the mass term in Eq.~\ref{eqn:lagrange}, resulting in a transition into the Haldane phase or the dimer phase, depending on the sign of the coupling constant. The other (marginally) relevant field, with dimension $\Delta_2=2$, governs the transition into the gapped NNN-AKLT phase (purple flow line in Fig.~\ref{fig:flow}).

  Note that the central charge $c=2$ suggests that the SU(2)$_4$ theory can be recast in the form of the 2 copies of bosonic fields. This is indeed possible~\cite{ZF.1986}, however these bosonic fields are highly non-local, explaining the non-trivial fractional scaling dimensions of the primary fields.

To summarize, we have demonstrated that the phase diagram of the spin-1 chain, shown schematically in Fig.~\ref{fig:flow}, is characterized by a continuous transition from the Haldane to the dimer phase for $-1\leq \beta < \beta^\ast\!\approx\! -0.2$, with vanishing spectral gap and divergent correlation length along the critical line (see Fig.~\ref{fig:xi-gap2}). For larger $\beta$, the transition becomes first-order, with the  spectral gap remaining open as $\alpha$ increases (see Fig.~\ref{fig:xi-gap1}), in agreement with previous DMRG~\cite{Kolezhuk.1996, Kolezhuk.1997-II} and field-theoretical studies~\cite{Allen.1995} at $\beta=0$. 
 We conjecture that separating these two regimes is a multicritical end-point at $\Omega=(\beta^\ast,\alpha^\ast)$, which terminates the line of second-order phase transitions. Everywhere to the left of this point, the RG flow along the critical line is governed by the marginally irrelevant perturbation to the SU$(2)_2$ WZW theory. However the point $\Omega$ itself lies in a different universality class of the SU$(2)_4$ WZW theory.

\subsubsection{Numerical results for the central charge}
\label{sec:central-charge}
\begin{figure}[t!]
\includegraphics[width=3.in]{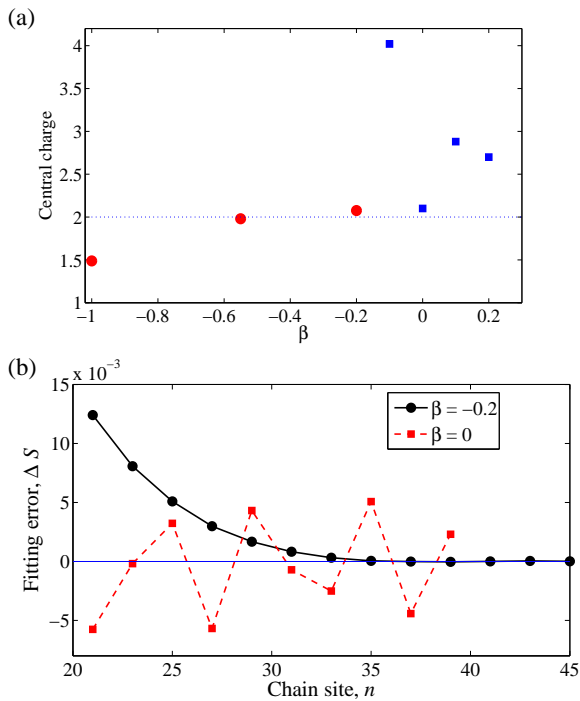}
\caption{(Color online) (a) Central charge $c$ for different values of $\beta$ along the critical line in Fig.~\ref{fig:flow}, extracted by fitting the DMRG entanglement entropy to Eq.~(\ref{eqn:entropy}) for $M=200$ kept states. The red circles mark the values of $c$ determined reliably, and the blue squares denote the data points where conformal scaling does not apply, as illustrated in (b): the error of fitting DMRG data to  Eq.~(\ref{eqn:entropy}) for $\beta=-0.2$ (reliable) and $\beta=0$ (scaling fails).}
\label{fig:central}
\end{figure}

To verify the field-theoretical predictions, we have extracted the central charge from our DMRG calculations of the entanglement entropy, known to scale in an open 1D system as follows~\cite{Holzhey.1994, Calabrese.2004}:
\beq
S(n) = S_0 + \frac{c}{6}\ln\left[\frac{2L}{\pi}\sin\left(\frac{\pi n}{L}\right)\right], \label{eqn:entropy}
\eeq
where $n$ is the site number in the chain of length $L$ marking the end of a contiguous block of $n$ sites for which the entanglement entropy is calculated. The central charge $c$ determined in this way is plotted in Fig.~\ref{fig:central}a for different values of $(\beta_c,\alpha_c)$ along the critical line in Fig.~\ref{fig:flow}. As mentioned earlier, the critical line is terminated at the critical end-point $\Omega=(\beta^\ast, \alpha^\ast)$.
We find that the entanglement entropy follows the scaling in Eq.~(\ref{eqn:entropy}) for values of $-1 \leq \beta_c<\beta^\ast$ and fails for larger values of $\beta_c$, as evident from the fitting error in Fig.~\ref{fig:central}b. This failure occurs because the system is no longer critical for $\beta_c>\beta^\ast$  and has a finite spectral gap, so the blue data points in Fig.~\ref{fig:central}a no longer have a meaning of a central charge.
From this, as well as from the DMRG calculation of the spectral gap, we were able to put brackets on the value of $\beta^\ast$ to lie in the interval $-0.2<\beta^\ast<-0.15$, with the corresponding bracket on $\alpha^*$ within $0.47<\alpha^*<0.53$.

At the TB point $\beta=-1$ we find the central charge to be $c=1.492$, very close to the theoretically expected value of $3/2$ for the SU$(2)_2$ WZW conformal theory. At $\beta=-0.2$, in the immediate vicinity of the multicritical point $\Omega$, we find a central charge of $2.08$, close to the value $c=2$ expected of the SU$(2)_4$ WZW theory that we propose in this study. In between these two points, the RG flow is expected to be governed by the attractive flow towards the TB point, so that in principle, one would expect the central charge to be $c=3/2$ everywhere along the critical line except at the multicritical point itself. However, the scaling analysis of the entanglement entropy from DMRG results in a value of central charge 1.98 at $\beta=-0.55$. This is likely because our analysis neglects logarithmic corrections of the marginally irrelevant operator ($\lambda$ in Eq.~\ref{eqn:lagrange}) to the finite-size spectrum and conformal scaling, first pointed out by Affleck \emph{et al}.\cite{Affleck.1989} Neglecting these logarithmic corrections in finite-size scaling analysis is known to sometimes lead to misleading results\cite{Affleck.1990}. Unfortunately, to the best of our knowledge, the effects of these logarithmic corrections on the entanglement entropy have not been worked out, and may well be responsible for the larger deduced values of central charge when using Eq.~\ref{eqn:entropy}. We would also like to point out another technical difficulty, in determining the precise position of the critical value $\alpha_c$ for a given value of $\beta$ in the phase diagram. Unlike the TB point whose position is known exactly, we relied on the maximum in the correlation length $\xi$ and minimum of the spectral gap $\Delta$ to determine the position of $\alpha_c(\beta)$, as illustrated in Figs.~\ref{fig:xi-gap2} and \ref{fig:xi-gap1}. Therefore, numerical errors in these quantities due to finite size effects may well have resulted in slightly inaccurate values of $\alpha_c$, which in turn would have affected the deduced values of central charge along the critical line. 

In summary, we find that the DMRG calculation of the central charge in the vicinity of the multicritical point $\Omega$ agrees well with the conformal field theory predictions for the SU$(2)_4$ WZW model, thus corroborating the field-theoretical analysis presented above.

\subsection{Edge Excitations}\label{sec:edge}
We now consider the nature of the ground state wavefunction as we move across the critical line.
As mentioned in the introduction, the Haldane phase (also referred to as AKLT phase) possesses effectively free $S=1/2$ spins on the edges with zero-energy edge excitations, which  give rise to the four-fold ($Z_2\times Z_2$) degenerate ground state for a finite chain with  open boundary conditions~\cite{Affleck.1987,Kennedy.1990}. As has been realized early on, the existence of these edge excitations is a hallmark of the topological nature of the  Haldane phase. By contrast, the dimer phase is not topological and lacks zero-energy edge excitations~\cite{Kennedy.1992, Xian.1993}. 
 The nature of the ground state wavefunction can thus be characterized by the existence or absence of gapless edge excitations.  

\begin{figure}[t!]
\includegraphics[width=3.in]{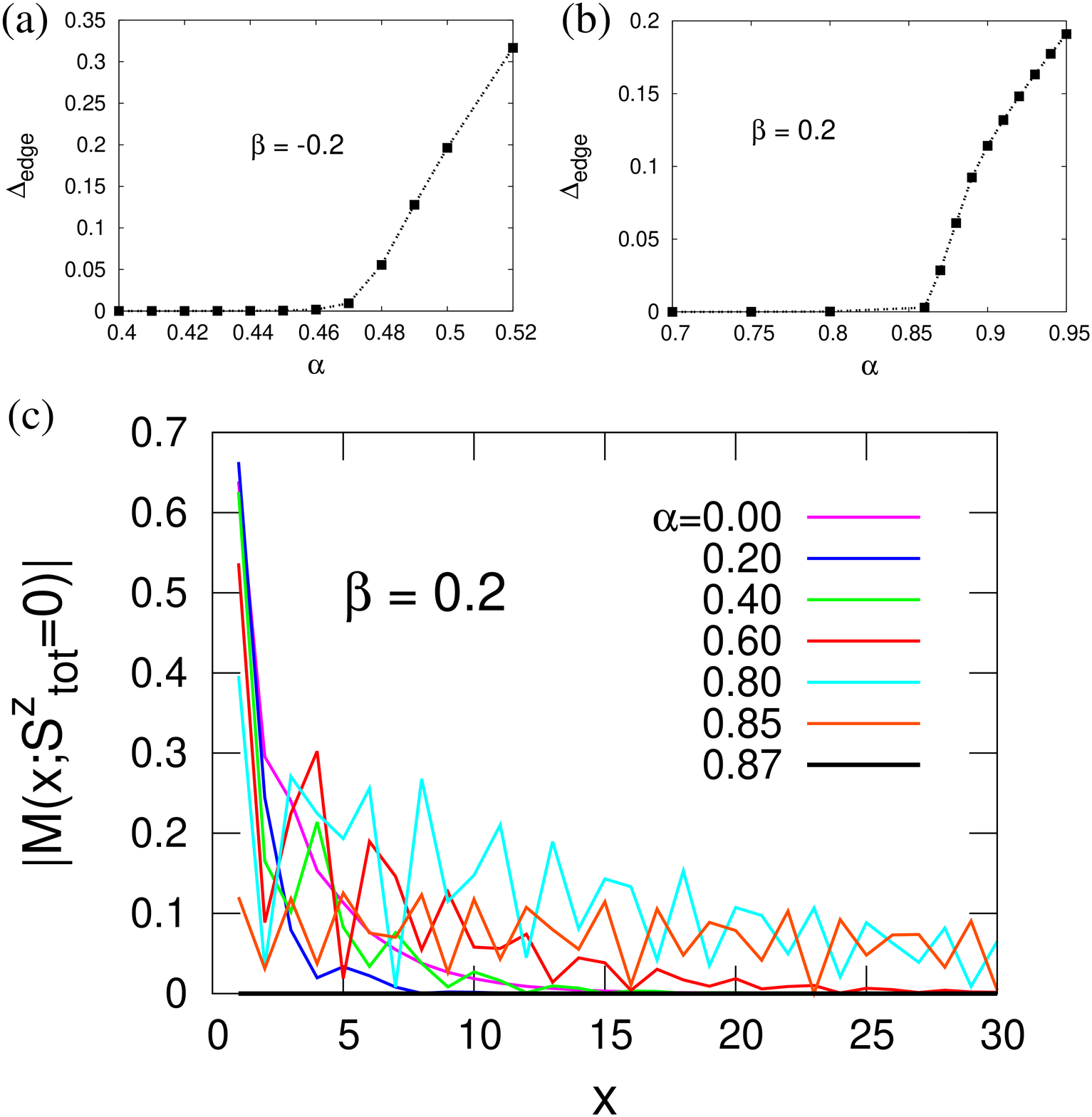}
\caption{(Color online)  The gap to edge excitations $\Delta_{\mathrm{edge}}$ at a fixed system size $L$ and fixed number of states $M=200$, opening up as $\alpha$ crosses the critical line from the AKLT phase (a) at $\alpha_c(\beta=-0.2)=0.480(5)$, $L=80$ and and (b) $\alpha_T(\beta=0.2)=0.855(5)$,  $L=100$.  We clearly see the edge excitations become gapped out.  (c) The absolute value of magnetization along the chain in the $S_{tot}=0$ symmetry sector approaching the critical line from the AKLT phase for $\beta=0.2$. 
We see the edge excitations begin to bleed into the bulk of the chain as the correlation length increases and then vanish as we cross the transition line.  
}
\label{fig:Edge-gap}
\end{figure}

Using the DMRG we are able to probe the edge excitations directly, by considering the magnetization $\langle  S^z(x) \rangle$ along the chain.  Edge excitations show up clearly in the Haldane phase in the form of a large magnetization confined to the chain ends~\cite{Kolezhuk.1997-II}, as shown in Fig.~\ref{fig:Edge-gap}c.   
Another way to probe the existence of the edge excitations in DMRG is by measuring a spectral gap between projections onto different total spin $S^z_{tot}$ sectors.  In the Haldane (AKLT) phase, the $S_{tot}=0$ ground state is degenerate with the first excited triplet in the $S_{tot}=1$ sector (the so-called ``Kennedy triplet''\cite{Kennedy.1990}), resulting in the aforementioned four-fold degenerate ground state in an open chain, which is the consequence of the so-called $Z_2\times Z_2$ spontaneous symmetry breaking~\cite{Kennedy.1992}. 
The lowest true bulk excitation lies in the  $S_{tot}=2$ sector~\cite{Kolezhuk.1996,Kolezhuk.1997-II}, and therefore, the bulk Haldane gap $\Delta$ is determined by the difference of the ground state energy in the symmetry sectors $S_{tot}=0$ and $S_{tot}=2$, and is plotted in Figures~\ref{fig:xi-gap2} and~\ref{fig:xi-gap1}. By contrast, the gap to edge excitations $\Delta_{\mathrm{edge}}$, which we emphasize is \emph{not} the true bulk gap, is the difference of the ground state energies between the symmetry sectors $S_{tot}=0$ and $1$. Therefore, the signature of the Haldane phase is the vanishing gap to edge excitations $\Delta_{\mathrm{edge}}=0$. 

Upon the transition to the translational-symmetry breaking 
dimer phase for $\alpha>\alpha_c(\beta)$ or to the NNN-AKLT phase for  $\alpha>\alpha_T(\beta)$,  the edge excitations become gapped out.  This is clearly seen in the gap $\Delta_{\mathrm{edge}}$ for both negative and positive $\beta$, as shown in Figures~\ref{fig:Edge-gap}a,b at a fixed system size.
  In addition, this transition manifests itself in the character of edge excitations, which we extract by plotting the magnetization along the chain in the ground state symmetry sector $M(x)|_{S_{tot}=0}=\langle S^z(x) \rangle$. On approaching the critical line from the Haldane phase below, the edge excitations bleed into the bulk of the chain, as shown clearly in Figure~\ref{fig:Edge-gap}c.  We therefore conclude that the nature of the ground state wavefunction changes from topologically non-trivial inside the Haldane phase to topologically trivial in both the dimer and NNN-AKLT phases above $\alpha_c$ and $\alpha_T$ respectively. 
  This conclusion is in accord with a recent work by  Gu and Wen~\cite{Gu.2009} who demonstrate that the Haldane phase for spin-1 chain is an example of the symmetry-protected topological phase (the symmetries are time-reversal, parity, and translational invariance).
This result was generalized to the case of odd-integer spin chains ($S=1,3,5\ldots$) by Pollmann \emph{et al.} in Ref.~[\onlinecite{Pollmann.2012}], who also showed that the dimer state is, by contrast, topologically trivial.

\section{Phases}
\label{sec:phases}
We now proceed to determine the topology of the phase diagram shown schematically in Figure~\ref{fig:schem-pd}.  By using a combination of the DMRG and an analytic MPS variational wavefunction we are able to identify and describe the nature of each phase as well as the character of the ground state wavefunction.  

\subsection{Haldane Phase}
 Fixing the value of $\beta$ and tuning $\alpha$ we find the Haldane phase (originally defined in the range $-1<\beta<1$ for $\alpha=0$) now extends over a region of finite $(\beta, \alpha)$ as shown in the phase diagram in Figure~\ref{fig:schem-pd}. 
 The Haldane (AKLT) phase is gapped and characterized by short range antiferromagnetic correlations.  The  spin-spin correlation function is well described by the OZ form in two dimensions (see equation~\ref{eqn:OZ} and section~\ref{sec:disorder} for discussion).  The AKLT phase also possesses a four-fold degenerate ground state and resulting gapless edge excitations, as discussed earlier in Section~\ref{sec:edge}.  Lastly, the AKLT phase is known to break a hidden $Z_2\cross Z_2$ symmetry and as a consequence the string order parameter is finite~\cite{Schollwock.1996}.  

\begin{figure}[ht!]
\centering
\includegraphics[width=2.75in]{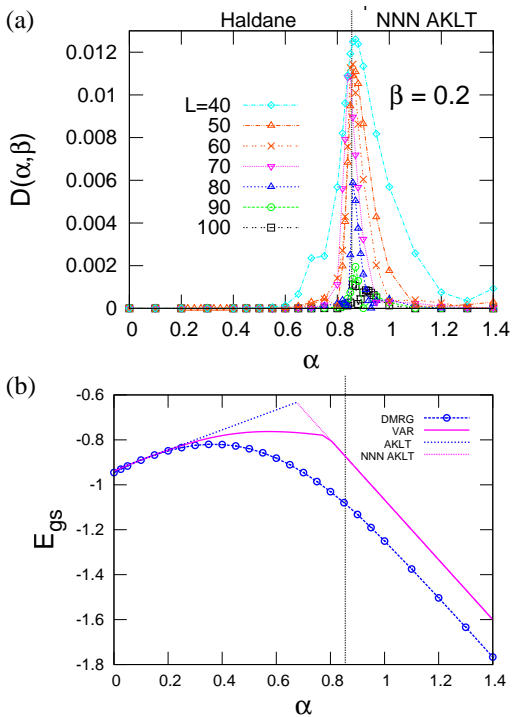}
\caption{(Color online) (a) The dimer order parameter $D(\alpha,\beta)$ as a function of $\alpha$ for $\beta=0.2$ for various system sizes $L$ and $M=200$ states. The dimer order parameter is identically zero in the Haldane phase, whereas inside the NNN-AKLT phase, we find that $D(\alpha,\beta)$ does not saturate in $L$ and may become vanishingly small in the thermodynamic limit. 
(b) Comparison of the ground state energy for $\beta=0.2$ as a function of $\alpha$ between the DMRG results extrapolated in $L$ and $M$ (circles), the AKLT ansatz (dashed line) and the variational wavefunction (continuous line).  For large values of $\alpha$ the ground state approaches the NNN-AKLT ansatz. The first order transition shows up clearly in both the AKLT and variational solutions, but the first numerical derivative of the ground state energy obtained within DMRG show no sign of a discontinuity at $\alpha_T$, consistent with references~[\onlinecite{Kolezhuk.1996, Kolezhuk.1997-II}].}
\label{fig:dimer-op}
\end{figure}

 We find the ground state wavefunction can be qualitatively described by the AKLT wavefunction ansatz which yields a ground state energy $E_{AKLT}(\alpha,\beta)=-4/3+4\alpha/9+2\beta$.  When compared to DMRG calculations, the naive estimate does quite a good job, while the variational estimate is more accurate, see Figures~\ref{fig:energy2}b and~\ref{fig:dimer-op}b.  We note that the AKLT wavefunction ansatz is almost exact along the disorder line and we will return to this point in detail in section~\ref{sec:results}.  We also find the four-fold degenerate ground state to survive up to the critical line as shown in Figure~\ref{fig:Edge-gap}.  Lastly, we have not calculated the string order parameter due to the limitations of the open source POWDER DMRG code~\cite{Rizzi.2008} used in this study, however based on the evidence of edge states, we expect the string order parameter to remain finite over the entire Haldane phase.

\subsection{Spontaneously Dimerized Phase} 
\label{sec:dimer}
\begin{figure}[t!]
\includegraphics[width=3.in]{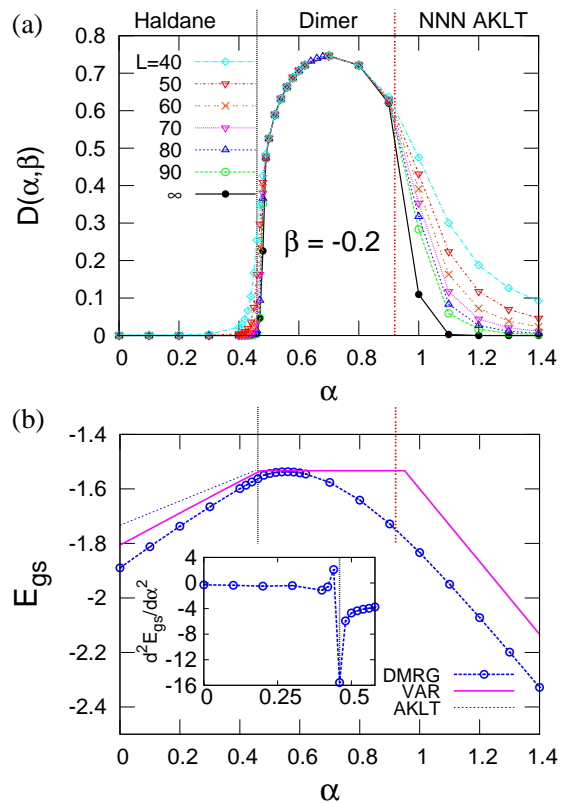}
\caption{(Color online) (a) The dimer order parameter $D(\alpha,\beta)$ as a function of $\alpha$ for $\beta=-0.2$ for various system sizes $L$ and $M=200$ states.  In the dimer phase we find the order parameter is independent of system size implying $D(\alpha,\beta)$ is finite in the thermodynamic limit, whereas deep inside the NNN-AKLT phase ($\alpha\gtrsim1.2$), we find that $D(\alpha,\beta)$ does not saturate in $L$ and likely becomes vanishingly small in the thermodynamic limit.
 (b) The ground state energy as a function of $\alpha$ for $\beta=-0.2$ obtained within DMRG extrapolated in $L$ and $M$ (circles) compared to the AKLT, NNN-AKLT, dimer, and variational wavefunctions.  Note, the naive estimate of the ground state energy of the dimer and NNN-AKLT wavefunctions agrees with the variational result and is therefore not shown.  We have clearly marked the critical line $\alpha_c$ as well as the first-order transition line $\alpha_{\delta}$ from the dimer to the NNN-AKLT phase (see Fig.~\ref{fig:schem-pd}).  Inset: The numerical second derivative of the ground state energy as a function of $\alpha$, we find a discontinuity in $d^2E_{gs}/d\alpha^2$ at the critical point, which suggests the transition into the dimer phase is second order.}
\label{fig:energy2}
\end{figure}

We now turn our attention to the dimerized phase.  
In order to characterize the amount of dimerization we find it useful to define the dimer order parameter  in the center of the chain:
\begin{equation}
D(\alpha,\beta)= | \langle {\bf S}_{L/2}\cdot{\bf S}_{L/2+1} - {\bf S}_{L/2-1}\cdot{\bf S}_{L/2} \rangle |,
\label{eqn:dimer}
\end{equation}
where the absolute value is necessary to account for the two different possible dimer coverings of the open chain.  We find that the dimer order parameter rises continuously from zero on entry into the dimer phase from the Haldane phase.
The dimerized phase is gapped with a ground state wavefunction well described by the dimerized wavefunction ansatz (see section~\ref{sec:methods}) as shown in Figure \ref{fig:energy2}b.  We show in section~\ref{sec:disorder} that the spin-spin correlation function is well described by a dimerized OZ form (see Eq.~\ref{eqn:OZD}).  Lastly, we find that the edge excitations are gapped in the dimerized phase (see Figure~\ref{fig:Edge-gap}a), as discussed earlier in section~\ref{sec:edge}.  

\begin{figure}[t!]
\includegraphics[width=3.in]{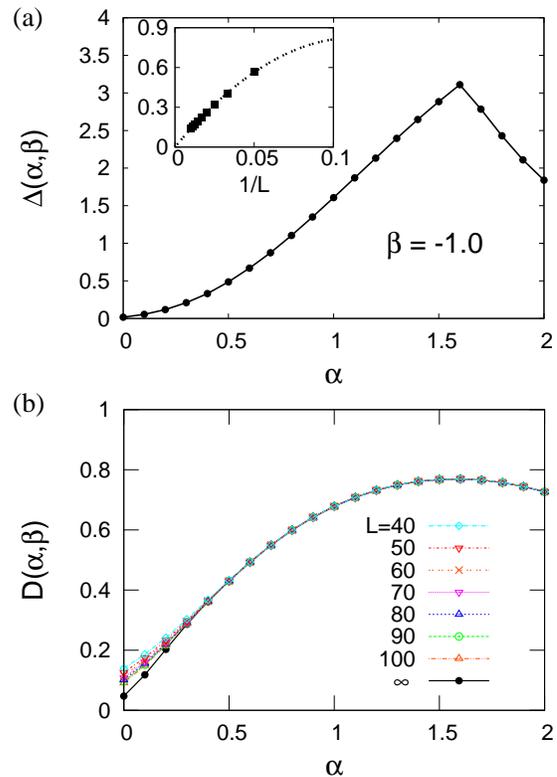}
\caption{(Color online) The bulk gap $\Delta$ between symmetry sectors $S_{tot}=0$ and $2$ for fixed $\beta=-1$ tuning away from the TB point extrapolated in $L$ with $M=200$ kept states (a).  We see the gap opens continuously for a finite value of $\alpha$.  Inset: Extrapolating the gap at the TB point ($\beta=-1.0$ and $\alpha=0$) to the thermodynamic limit, the fit is in excellent agreement with a gapless point namely, $\Delta(L) = a_{\Delta}/L + b_{\Delta}/L^2$.  When fit to a functional form for a finite gap [see equation (\ref{eqn:thermo-limit})] we find a small gap at the TB equal to $0.017(2)$.  (b) Dimer order parameter, defined in equation~\ref{eqn:dimer} as a function of the system size $L$ and extrapolated to the thermodynamic limit for $M=200$ states, clearly displaying the system enters the dimer phase immediately upon tuning $\alpha$ away from the TB point.  The finite value of $D(\alpha=0,\beta=-1)$ at the TB point is attributed to not reaching large enough system sizes at the critical point.}
\label{fig:gap-TB}
\end{figure}

\begin{figure}[t!]
\includegraphics[width=2.25in, angle= -90]{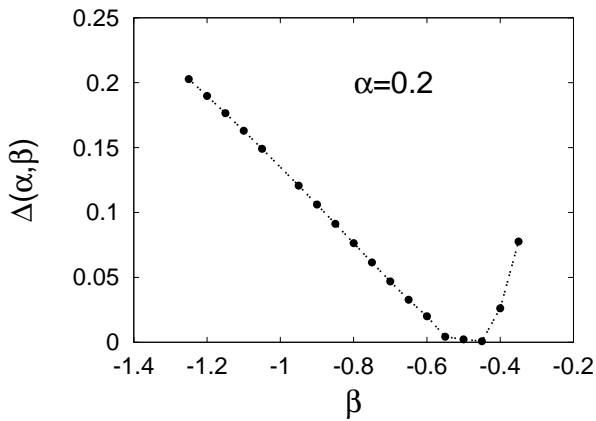}
\caption{Bulk gap $\Delta$ in the vicinity of the TB point for a fixed NNN coupling $\alpha = 0.2$, as a function of $\beta$ extrapolated in system size $L$ with $M=200$ kept states. From the behavior of the gap, it is clear the critical line will move towards a smaller value of $\beta$, as we increase $\alpha$ away from the TB point, verifying the slope of $\alpha_c(\beta)$ shown in the schematic phase diagram~\ref{fig:schem-pd}. }
\label{fig:gap-v-beta}
\end{figure}

In order to determine the boundaries of the dimerized phase, we first consider tuning away from the TB point with a finite $\alpha$, (i.e. fixing $\beta=-1$).  With the DMRG at the TB point we find a ground state energy $E_{\mathrm{GS}}=-3.999(1)$ in agreement with the exact Bethe ansatz result~\cite{Takhtajan.1982,Babudjian.1982} of $-4.0$.  In addition, as previously discussed, the TB point has a central charge of $c=1.5$ while we obtain $c=1.49$ with the DMRG.  Within the numerical accuracy, we find a finite bulk gap $\Delta(\alpha,\beta)$ which opens up for infinitesimal $\alpha>0$, see Figure~\ref{fig:gap-TB}.  In addition, we find that the dimer order parameter grows continuously upon moving away from the TB point, clearly marking the entry into the dimerized phase.  We remark that in the vicinity of the TB point, the diverging correlation length makes the identification of the critical line $\alpha_c(\beta)$ quite challenging without going to much larger chain sizes than we have in the present study.

In order to determine the slope of the critical near the TB point ($\beta=-1$), we consider fixing the NNN coupling $\alpha=0.2$ and varying $\beta$ to determine if the transition moves to either a larger or smaller value of $\beta$.  By calculating the gap, shown in Figure \ref{fig:gap-v-beta}, it is quite clear that the critical line moves right towards a smaller value of $|\beta|$.  We find a  small gap [$\Delta(\alpha=0.2,\beta=-0.5) = 0.0022$] at the critical point $\beta_c(\alpha=0.2) \approx -0.50(5)$.
Interestingly, as shown in Figure~\ref{fig:gap-v-beta} for $\beta<\beta_c$, we find that the gap scales linearly with $\beta$: $\Delta(\alpha,\beta) \propto |\beta_c(\alpha) - \beta|$, similar to the behavior in the vicinity of the TB point.

We now consider a range of parameters by fixing $\beta$ in the range $-1.0<\beta\leq -0.2$ and tuning $\alpha$.  In the following, we present results for $\beta=-0.2$ which we find to be close to the proposed multi-critcal point $\Omega$.  The dimer order parameter in the thermodynamic limit (see Figure~\ref{fig:energy2}a) first grows continuously upon entering the dimer phase and then decreases on entering the NNN-AKLT region of the phase diagram.  We find the ground state energy to agree very well with the analytic result from the dimerized wavefunction MPS ansatz which yields an $\alpha$-independent ground state energy $E_{Dimer}(\alpha,\beta)=8\beta/3 -1$ (see section~\ref{sec:MPS} for more detail).  In addition, DMRG finds a discontinuous second derivative of the ground state energy as we cross the critical line $\alpha_c(\beta)$ (see Fig.~\ref{fig:energy2}b), implying that the phase transition is of second order, consistent with the field theoretical discussion in section~\ref{sec:critical-line}, the very small bulk DMRG gap (see Fig.~\ref{fig:xi-gap2}a), and the diverging correlation length at the transition (Fig.~\ref{fig:xi-gap2}b).  

Interestingly, in the vicinity of $-0.2 < \beta^\ast < -0.15$ and $0.47<\alpha^\ast<0.53$, we find the critical and disorder lines intersect (see section~\ref{sec:disorder} for a discussion of disorder points).  This has serious physical implications, since the correlation length $\xi$ goes through a \emph{minimum} at the disorder point, meaning that it is not possible for $\xi$ to diverge had this been a $2^{\text{nd}}$ order phase transition.  Therefore, consistent with the field theoretical results, the critical line $\alpha_c(\beta)$ must terminate at the multicritical point $\Omega(\alpha^\ast,\beta^\ast)$, merging into a first-order transition line $\alpha_T$ as shown in the phase diagram Fig.~\ref{fig:schem-pd}.  For $\beta=-0.175$ to the right of the multicritical point, the correlation length at the transition is only moderately enhanced on the order of $\xi(\alpha,\beta)=10.0$ (in units of the lattice spacing) and the edge excitations become gapped.  For the case of $\beta=-0.15$, being in close vicinity of the disorder line makes the identification of $\alpha_T(\beta)$ from correlation length difficult. Instead, we track the transition via the suppression of gapless edge excitations (see section~\ref{sec:edge}), resulting in the value $\alpha_T(\beta=-0.15)= 0.530(5)$ which actually agrees well with the finding from the analytical MPS wavefunction ansatz (dashed line in Fig.~\ref{fig:schem-pd}).

\subsection{NNN-AKLT}
\label{sec:NNN}
\begin{figure}[h!]
\includegraphics[width=3.in]{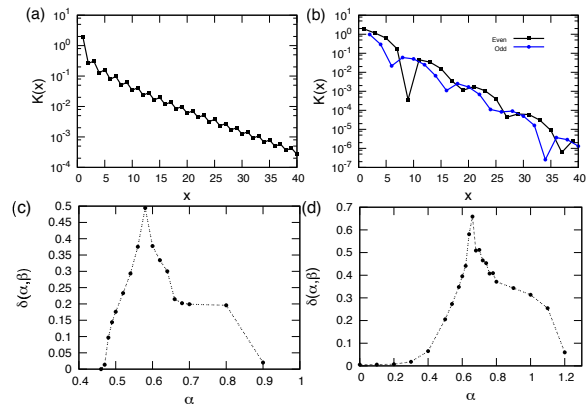}
\caption{(Color online)  The real space correlation function $C(x)$, plotted in terms of $K(x)\equiv C(x)(-1)^x\sqrt{x}$ for $\beta=-0.2$ with: (a) $\alpha=0.50$ in the dimer phase with
$q(\alpha,\beta)=\pi$; and (b) $\alpha=0.80$ in the incommensurate dimer
phase, plotted for even (black) and odd (blue) values of $x$ to show
clearly finite dimerization on top of an OZ form with $q(\alpha,
\beta)>\pi$.
The dimerization $\delta(\alpha,\beta)$ is plotted  as a function of
$\alpha$ for (c) $\beta=-0.2$ and (d) $\beta=-0.3$. We see that the
dimerization first rises continuously as we cross $\alpha_{\delta}^l(\beta)$, before acquring a cusp at the C-IC transition ($\alpha_{dL}$ line), and then decreases to zero in the incommensurate dimer phase.   The transition line $\alpha_{\delta}(\beta)$ between the incommensurate dimer phase and the NNN-AKLT phase is defined when dimerization $\delta(\alpha,\beta)$ becomes vanishingly small in Eq.~(\ref{eqn:OZD}).}
\label{fig:DO}
\end{figure}

The NNN-AKLT phase~\cite{Kolezhuk.1996,Kolezhuk.1997-II} is gapped with incommensurate spin-spin correlations in both real and momentum space, $q\neq \pi$ in Eq.~(\ref{eqn:OZ}).
Unlike the dimer phase, we find that the NNN-AKLT phase does not break the translational symmetry of the lattice, manifested by the fact that the calculated dimer order parameter $D$ in Eq.~(\ref{eqn:dimer}) becomes vanishingly small in the thermodynamic limit $L\to\infty$, as illustrated in  Figures~\ref{fig:energy2}a and \ref{fig:dimer-op}a. In addition,  the dimerization $\delta(\alpha,\beta)$ in the spin-spin correlation function Eq.~(\ref{eqn:OZD}) also vanishes, see Figure~\ref{fig:DO}.  

Because of this symmetry distinction between the dimer and NNN-AKLT phases, there must be a phase transition between the two, with the NNN-AKLT phase stable above $\alpha_\delta$ line in the phase diagram Fig.~\ref{fig:schem-pd}. Based on the fact that the spectral gap never closes  and the correlation length remains finite at $\alpha_\delta$, we conclude that the transition must be first order. Numerically, the location of the $\alpha_\delta(\beta)$ line is determined to be where $\delta(\alpha,\beta)\rightarrow 0$ in the spin-spin correlation function Eq.~(\ref{eqn:OZD}). Alternatively, one can choose to determine the NNN-AKLT phase boundary from the condition that the dimer order parameter $D$ vanishes in the thermodynamic limit, which turns out to give a slightly larger value than $\alpha_\delta$. 
 In this work, we have chosen the former method of determining $\alpha_\delta$ as it gives more accurate results with a weaker system size dependence.

The NNN-AKLT phase is also distinct from the Haldane phase, as manifest by the absence of zero-energy edge excitations which become separated by a finite gap from the unique ground state. To see this explicitly, consider first the $\alpha\to \infty$ limit (i.e. infinitely strong NNN coupling), in which case the system decouples into two copies of  spin-$1$ chain, each with 4-fold degenerate ground state (the Kennedy triplet coincides with the singlet). Therefore, the ground state of two decoupled chains is $4\times4=16$ fold degenerate~\cite{Kolezhuk.1996,Kolezhuk.1997-II}. However, any finite $J_1$ is sufficient to couple the edge spins of the two chains, which then form pairwise singlets to result in a unique NNN-AKLT ground state, whose energy is lowered by the value of order of $J_1$. The remaining edge excitations are at higher energies. 
We thus use the appearance of a finite gap in the edge state spectrum as a signature of transition from the Haldane into the NNN-AKLT phase. In addition, we track the entry into the NNN-AKLT phase by a maximum in the spin-spin correlation length $\xi$ in Eq.~(\ref{eqn:OZ}) and the spectral gap passing through a minimum.~\cite{Kolezhuk.1997-II} We note that historically, the vanishing of the string order parameter was used\cite{Kolezhuk.1996,Kolezhuk.1997-II} to determine the transition into the NNN-AKLT phase above $\alpha>\alpha_T$. Because of the technical limitations of our DMRG code, we were unable to calculate the string order parameter, however the location of the transition $\alpha_T(\beta=0)$ that we determined as described above coincides with the value $\alpha_T\approx 0.74$ determined from the string order parameter in Ref.~\onlinecite{Kolezhuk.1997-II}.

We find that the NNN-AKLT phase exists over a wide range of $\beta$ provided $\alpha >\alpha_T$ is large enough.  In particular, we considered several values of $\beta$ in the range $[-0.125,0.4]$ and calculated scans along the $\alpha$ axis to determine the transition line $\alpha_T(\beta)$, shown in the phase diagram in Figure~\ref{fig:quant-pd}.  In addition, we find that the NNN-AKLT wavefunction is well described by the analytic matrix product state  ansatz (see section~\ref{sec:methods}).  This is shown in Figures~\ref{fig:energy2}b and~\ref{fig:dimer-op}b, where the ground state energy acquires the same slope as the analytic MPS result: $E_{NNN}(\alpha,\beta)=4/3(\beta-\alpha)$.

\section{Short Range Order}
\label{sec:disorder}
\subsection{Introduction to Disorder and Lifshitz Points}
As mentioned in the introduction, Haldane's mapping~\cite{Haldane.1983} to the two dimensional non-linear sigma model indicates that the antiferromagnetic quantum spin chain can be regarded as a two-dimensional classical spin model at a temperature $T_{\mathrm{eff}}\propto 1/S$.  As a consequence of the Mermin Wagner theorem~\cite{Mermin.1966},  the classical two-dimensional Heisenberg model with short-range interactions cannot break a continuous symmetry at a finite temperature.  Therefore, as a result of Haldane's mapping, the integer spin quantum Heisenberg  chain cannot exhibit long-range magnetic order.  However, it is possible to break a discrete translational symmetry, as is the case in the dimerized phase.   

Without breaking the continuous SU(2) spin symmetry, it is possible for the quantum spin chain to possess short range order as discussed earlier in section~\ref{sec:sr-phases}.  It has been shown that tuning either the biquadratic~\cite{Schollwock.1996}  or next nearest neighbor interaction~\cite{Kolezhuk.1996,Kolezhuk.1997-II} can introduce short range incommensurate order, whose onset occurs at the so-called disorder point or Lifshitz point characterizing incommensurate spin correlations in either real or momentum space, respectively.  

Disorder points of the first kind and Lifshitz points have been well defined in classical statistical mechanics~\cite{Stephenson} and have been discussed extensively in the context of quantum spin-1 chains in Refs.~\onlinecite{Schollwock.1996,Kolezhuk.1996,Kolezhuk.1997-II} (and references therein). We therefore only briefly review these concepts here.  There are two types of disorder points: of the first and second kind~\cite{Stephenson, Schollwock.1996}, and in this work we shall only encounter a disorder point of the first kind, so we shall refer to it simply as a \emph{disorder point} in what follows.  At this point, the real space spin correlation function acquires an incommensurate  Ornstein--Zernicke form  with a wave-vector $q\neq \pi$  in Eq.~(\ref{eqn:OZ}).

\begin{figure}[t!]
\includegraphics[width=3.in]{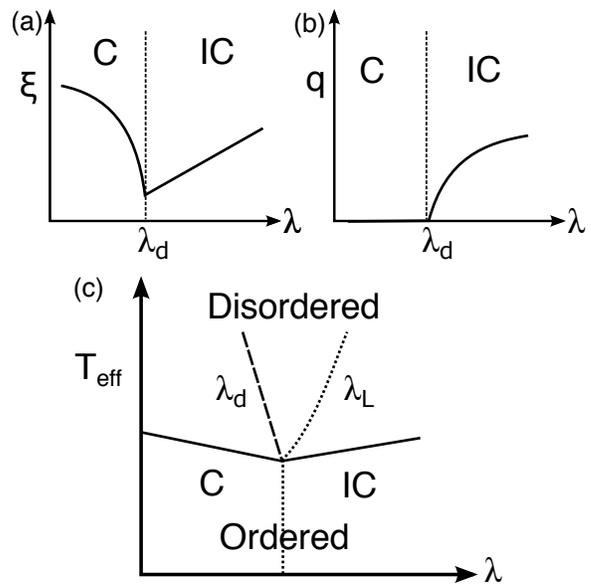}
\caption{Schematic figures displaying a disorder point in the correlation length $\xi$ (a) and the wave number $q$ (b).  C and IC denote commensurate and incommensurate correlations in real space.  Generic phase diagram of breaking a discrete symmetry via tuning an effective energy scale $T_{\rm eff}$ in two dimensions in the presence of incommensurate correlations $\lambda$ (c). 
\label{fig:xiandq}}
\end{figure}

Tuning the control parameter $\lambda$ of a Hamiltonian across the phase diagram, the system will pass through a disorder point at $\lambda_d$ if the correlation length $\xi_A(\lambda)$ 
 develops an infinite slope on the commensurate side but is generally finite on the incommensurate side, i.e. $\ud\xi_C(\lambda_d)/\ud\lambda=\infty$ and $\ud\xi_{IC}(\lambda_d)/\ud\lambda<\infty$.  In addition, the wave number of the correlation function $q_A(\lambda)$ changes from a commensurate to an incommensurate value at $\lambda_d$.  In the commensurate phase $q_C(\lambda<\lambda_d)$ is constant so that $\ud q_C(\lambda_d)/\ud\lambda = 0$, whereas 
on the incommensurate side the wave number rises continuously:
\begin{equation}
q_{IC}(\lambda)-q_C(\lambda_d) \propto (\lambda - \lambda_d)^{\sigma},
\label{eqn:qd}
\end{equation}
with a non-universal exponent $\sigma$.  The generic behavior of $\xi$ and $q$ across a disorder point are shown in Figures~\ref{fig:xiandq} (a) and (b).  Interestingly, these features have been found across numerous different physical scenarios in both classical and quantum models.  

Upon further tuning the control parameter $\lambda$, the system will pass through a Lifshitz point at $\lambda_L$ where the correlation function in momentum space goes from a single commensurate peak at $q=\pi$ to a two-peak incommensurate structure in Eq.~(\ref{eqn:S(q)}).
 In the disordered phase, the disorder and Lifthitz lines are distinct from each other. In a true broken symmetry state (for instance in a classical three-dimensional system below the magnetic ordering temperature), the disorder and Lifshitz transitions merge into a single line that separates the long-range commensurate from long-range incommensurate order, as shown schematically in Figure~\ref{fig:xiandq}(c).
  We would like to point out the remarkable similarity between this generic classical phase diagram and our results in Figure~\ref{fig:quant-pd}. Indeed, we find that in the short-range Haldane phase, the disorder and Lishitz transition lines are distinct from each other, forming the boundaries between the commensurate (C), real-space incommensurate (IC-R) and fully incommensurate (IC) short-range spin order. By contrast, these two lines merge into a single C/IC transition inside the symmetry-broken dimer phase, see Fig.~\ref{fig:quant-pd}.
 In the remainder of this section, we explain these findings in more detail, focusing in particular on the regime of positive $\beta$ (Haldane phase) and negative $\beta$ (dimerized phase).


\subsection{Results: $0<\beta<1$}
\label{sec:results}

In this subsection, we show that by varying $\alpha$ while keeping $\beta>0$,  the quantum spin-1 chain develops incommensurate short-range order inside the Haldane phase by passing first through a disorder transition and  then a Lifshitz transition. 

\subsubsection{Disorder and Lifshitz lines}
By fitting our DMRG results for the spin-spin correlation function $C(x)$ to the OZ form Eq.~(\ref{eqn:OZ}), we extract both the correlation length $\xi(\alpha,\beta)$ and the wave number $q(\alpha,\beta)$.
Fixing $\beta = 0.05, 0.10, 0.20, 0.30$, such that we remain to the left of the AKLT point~\cite{Affleck.1987} ($\beta = 1/3, \alpha=0$), we start from the commensurate phase with a wave vector $q=\pi$ in Eq.~(\ref{eqn:OZ}) and then tune $\alpha>0$ until we pass through a disorder transition at $\alpha_d(\beta)$. We find that the AKLT point itself lies on the disorder line, in agreement with the earlier DMRG work by Schollw\"ock and collaborators~\cite{Schollwock.1996}.

\begin{figure}[t!]
\includegraphics[width=3.in]{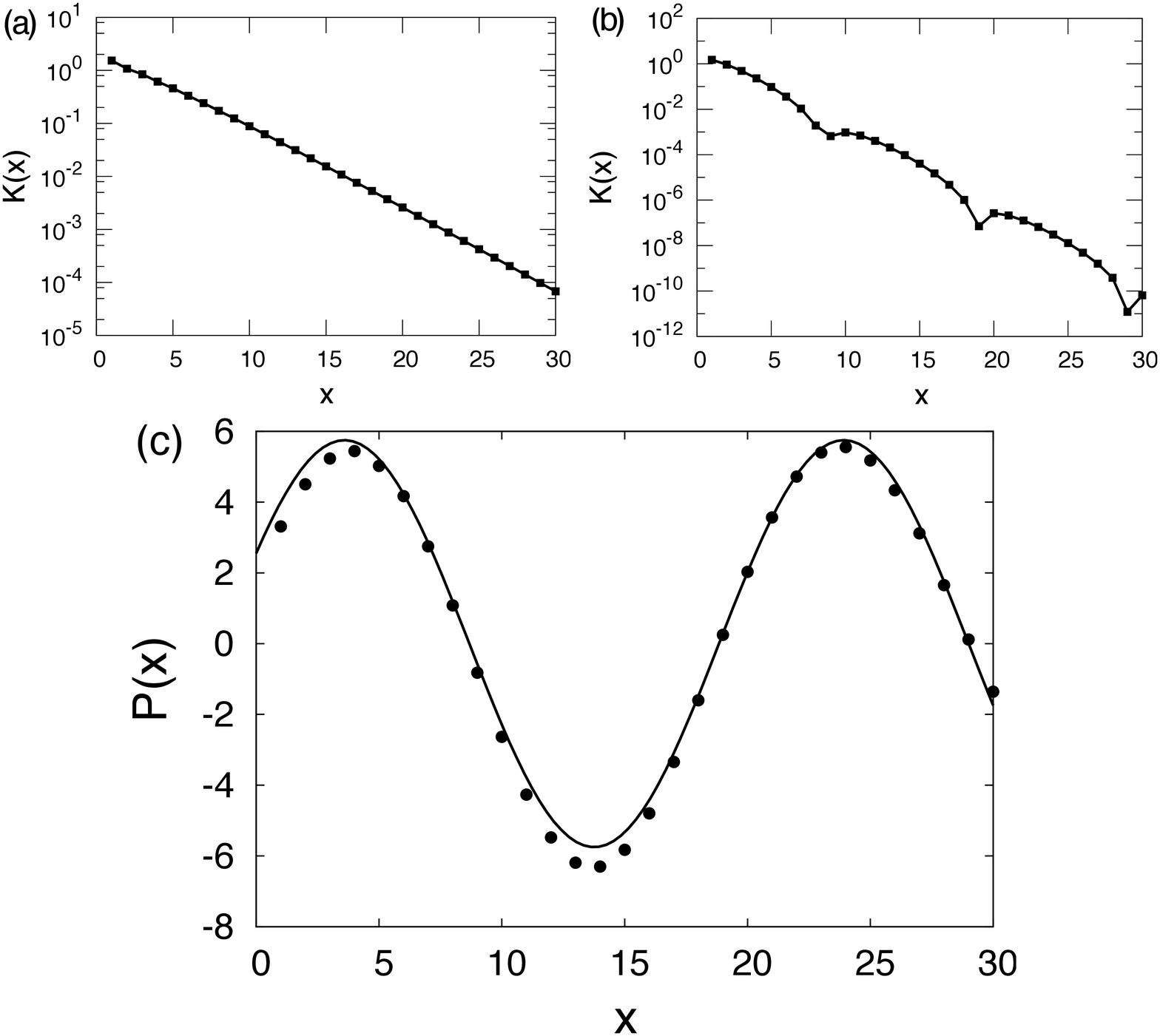}
\caption{The real space correlation function $C(x)$, plotted in terms of $K(x)\equiv C(x)(-1)^x\sqrt{x}$ to extract the correlation length $\xi(\alpha,\beta)$ for $\beta=0.1$ on either side of the disorder point at $\alpha_d(\beta=0.1)=0.180(2)$ with (a) $\alpha=0.1<\alpha_d(\beta)$ and  (b) $\alpha=0.2>\alpha_d(\beta)$.  Note the presence of incommensurate real space correlations in (b) in the form of peaks.  Extracting the wave number $q(\alpha,\beta)$ by plotting $C(x)$ in terms of $P(x)\equiv K(x)\exp(x/\xi)$ for $\beta=0.1$ and $\alpha=0.2$, the numerical data are circles and the solid line is a fit to the data (c).  }
\label{fig:chix}
\end{figure}

To extract the correlation length $\xi$, we fit the numerical data to $K(x)\equiv C_{OZ}(x)(-1)^x\sqrt{x} = \cos((q-\pi)\cdot x)e^{-x/\xi}$, using a procedure similar to that described in Ref.~\onlinecite{Schollwock.1996}. Namely, for $\alpha<\alpha_d(\beta)$, $q=\pi$ and we can directly fit our $K(x)$ to an exponential form, whereas for $\alpha>\alpha_d(\beta)$ we fit the maxima of the function $K(x)$.  
Once the correlation function $\xi$ has been determined, we extract the wave number $q(\alpha,\beta)$ by fitting the function $P(x)\equiv K(x)\exp(x/\xi)$ to the cosine form, see Figure~\ref{fig:chix}.  We find that the real space spin-spin correlation function in this region of the phase diagram is indeed well described by the two-dimensional Ornstein--Zernicke form in equation (\ref{eqn:OZ}).  

\begin{figure}[h!]
\includegraphics[width=3.in]{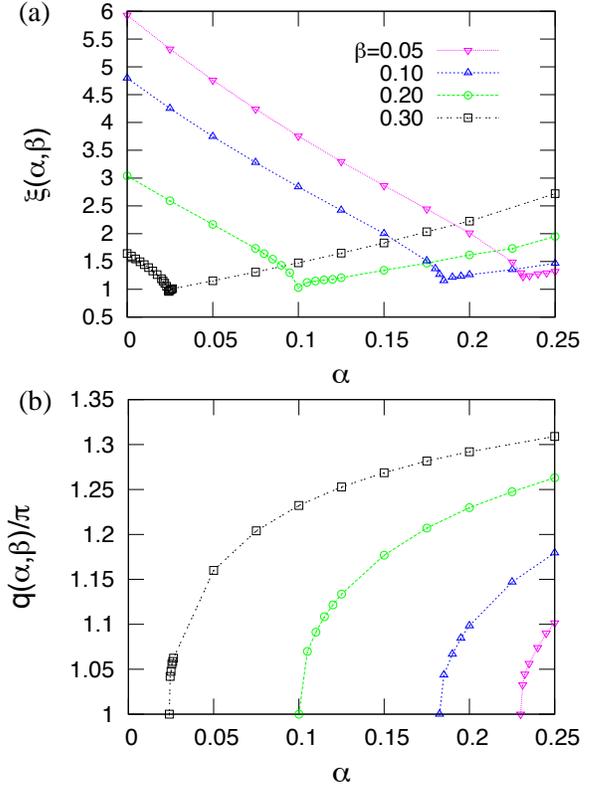}
\caption{(Color online)  (a) The correlation length $\xi(\alpha,\beta)$ for various values of $\alpha$ and $\beta$.  We identify the minimum in $\xi(\alpha,\beta)$ as a disorder point. Note that the value of the correlation length at the disorder point is increasing as $\beta$ decreases. (b) The wave number $q(\alpha,\beta)$ for fixed $\beta$ as a function of $\alpha$.  We find $q(\alpha,\beta)$ rises continuously from $\pi$ at $\alpha_d(\beta)$ to an incommensurate value.}
\label{fig:xq}
\end{figure}

In Figure~\ref{fig:xq}, we present the correlation length $\xi(\alpha,\beta)$ for various values of $\alpha$ and $\beta$.  We find that $\xi(\alpha,\beta)$ experiences a minimum at $\alpha_d(\beta)$ with a large slope for $\alpha<\alpha_d(\beta)$.  In addition, over the same set of $\alpha$ and $\beta$, we find the wave number $q(\alpha,\beta)$ to grow  continuously from $\pi$ for $\alpha>\alpha_d(\beta)$, see Figure~\ref{fig:xq}.  In the vicinity of $\alpha_d(\beta)$, we have determined the exponent $\sigma(\alpha,\beta)$ defined in equation (\ref{eqn:qd}) [with $\lambda$ replaced by $\alpha$ and $\lambda_d$ replaced with $\alpha_d(\beta)$]. In each case, the exponent satisfies the inequality $0<\sigma(\alpha,\beta) < 1$ consistent with $q(\alpha,\beta)$ having an infinite slope at $\alpha_d$ on the incommensurate side.  These results allow us to conclude that each $\alpha_d(\beta)$ is in fact a disorder point which taken together, define a line of disorder transitions in the $\beta-\alpha$ phase diagram, see Fig.~\ref{fig:quant-pd}.  Our results indicate that the disorder line smoothly connects the AKLT point in the biquadratic chain ($\alpha=0$) to the disorder point in the NNN chain ($\beta=0$) found in previous DMRG studies.\cite{Schollwock.1996,Kolezhuk.1996,Kolezhuk.1997-II}

\begin{figure}[h!]
\begin{minipage}{20pc}
\includegraphics[width=3.2in]{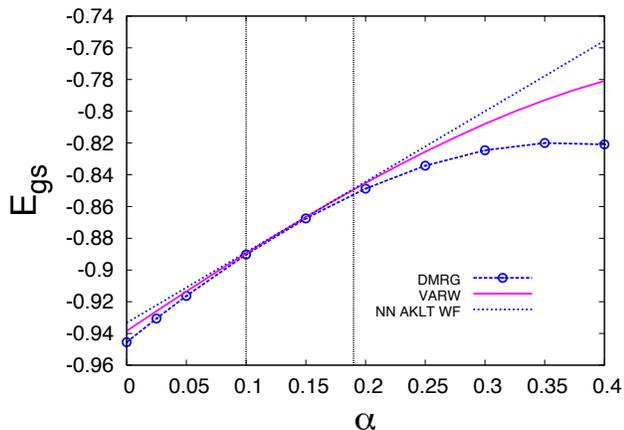}
\end{minipage}
\caption{(Color online)  Comparison of the ground state energy for $\beta=0.2$ as a function of $\alpha$ between the DMRG results (points), the AKLT ansatz (dashed line) and the variational wavefunction (continuous line).  Zoomed in region around the disorder and Lifshitz points (dashed lines), upon crossing the disorder point the ground state is very close to the AKLT ansatz.}
\label{fig:Egs-zoom}
\end{figure}

Precisely at the disorder line $\alpha_d(\beta)$, our results for the spin-spin correlation function can be fit with a purely exponential decay $C(x)\sim \cos(qx) e^{-x/\xi}$, corresponding to the one-dimensional Ornstein--Zernicke form, i.e.  $d=1$ in the Eq.~(\ref{eqn:OZ}), rather than the conventional $d=2$ that one expects from the quantum-to-classical mapping.   
As briefly mentioned in section~\ref{sec:sr-phases}, this behavior of dimensional reduction is expected to occur at the disorder point. In particular, this was shown to be the case at the AKLT point\cite{Schollwock.1996} where the identification is possible thanks to the exactly known ground state.\cite{Affleck.1987}  
Our results show that the same behavior is true along the entire disorder line $\alpha_d(\beta)$.  
Intriguingly, although the AKLT wavefunction ansatz is only approximate away from the AKLT point, we find that the entire region of the $\beta-\alpha$ phase diagram between disorder and Lifshitz lines, namely $\alpha_d(\beta)<\alpha(\beta)<\alpha_L(\beta)$ is very well described by the AKLT ground state, see Fig.~\ref{fig:Egs-zoom}. Physically, one can think of the disorder line as marking the entry into the AKLT ground state region for a range of $\alpha$.

\begin{figure}[h!]
\includegraphics[width=2.2in, angle= -90]{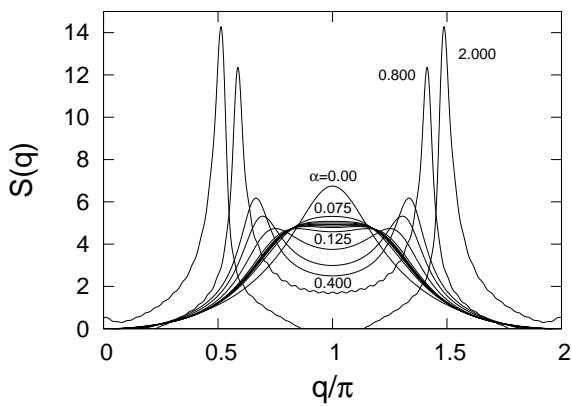}
\caption{The momentum space correlation function $S(q)$ for $\beta=0.3$ and various values of $\alpha$.  We find a Lifshitz point at $\alpha_L(\beta=0.3)=0.095(2)$. }
\label{fig:chiq}
\end{figure}

We now turn to the correlation function in momentum space $S(q)$, see Eq.~(\ref{eqn:S(q)}).
We consider $\beta =  0.05, 0.10, 0.20, 0.30, 0.35, 0.40$, such that the system remains to the left of the Lifshitz point for $\alpha=0$, which is known to lie at $\beta_L(\alpha=0)=0.43806(4)$ (see Ref.~\onlinecite{Bursill.1995}).  Tuning $\alpha>0$ we find a line of Lifshitz points above the disorder line, $\alpha_L(\beta)>\alpha_d(\beta)$, where the peak in $S(q)$ shifts from $q=\pi$ to an incommensurate double-peak structure, see Figure~\ref{fig:chiq}. For large $\alpha\to\infty$, the wave-vector saturates at $q=\pm \pi/2$, which is understood as a consequence of the doubling of the lattice spacing in the pure NNN chain.

\subsection{Results: $-1< \beta < 0$}
As we have shown previously (see section~\ref{sec:dimer}), the effect of a large negative $\beta$ is to form dimers between neighboring spins, so that even in the presence of a finite $\alpha$ the spin chain is spontaneously dimerized.   
When the system is inside the Haldane phase but close to the boundary $\alpha_c(\beta)$ with the dimer phase, it is possible for the spin-1 chain to experience a short-range dimer (SD) order, even though the translational symmetry is not broken in the thermodynamic limit. Our data indeed support the existence of such an SD phase in the small region $-0.2\lesssim\beta\lesssim -0.15$ close to the multicritical point $\Omega$ (see Fig.~\ref{fig:quant-pd}).
 Similar to the disorder transition discussed earlier, a signature of such short-range dimer phase will appear in the  real-space spin-spin correlation function, which will maintain the OZ form while acquiring an additional dimerization, $\delta(\alpha,\beta)>0$ in Eq.~(\ref{eqn:OZD}).

\subsubsection{Disorder, Lifshitz and Dimerization lines}
We define the lower and upper dimerization crossover lines as $\alpha_{\delta}^l(\beta)$ and $\alpha_{\delta}^u(\beta)$ respectively, where short-range dimerization $\delta(\alpha,\beta)$ is finite for $\alpha_{\delta}^l(\beta)<\alpha<\alpha_{\delta}^u(\beta)$. Above the upper bound  $\alpha>\alpha_{\delta}^u(\beta)$, the correlation function can be fitted using the standard non-dimerized OZ form, Eq.~(\ref{eqn:OZ}).
In the range $-0.125 \lesssim \beta<0$, fixing $\beta$ and tuning $\alpha$ makes the model pass through the disorder and Lifshitz transitions at $\alpha_d(\beta)<\alpha_L(\beta)<\alpha_T(\beta)$, until eventually the NNN-AKLT phase is reached above the first-order transition line $\alpha_T(\beta)$, see Figs.~\ref{fig:schem-pd} and \ref{fig:quant-pd}.
The disorder and Lifshitz transitions have the same properties as described in the previous subsection for $\beta>0$.  The only distinction is that due to a negative $\beta$, we find $\alpha_d(\beta)=\alpha_{\delta}^l(\beta)$, i.e. upon crossing the disorder line, we find a short range incommensurate dimerized phase, with both $q(\alpha,\beta)>\pi$ and $0<\delta(\alpha,\beta)<1$ in Eq.~(\ref{eqn:OZD}).  
When both $q(\alpha,\beta)$ and $\delta(\alpha,\beta)$ are small, it is difficult to accurately determine the precise value of the dimerization, however it is clear when the standard OZ form in Eq.~(\ref{eqn:OZ}) is a good fit, and we use this to determine $\alpha_{\delta}^u(\beta)$, which in this regime always lies below the Lifshitz transition $\alpha_{\delta}^u(\beta)<\alpha_L(\beta)$, see Fig.~\ref{fig:quant-pd}.

As we have discussed previously, for the range $\beta^\ast  \lesssim \beta < 1$ the first-order transition line into the NNN-AKLT phase $\alpha_T(\beta)$ lies above the disorder and Lifshitz lines in the $\beta-\alpha$ phase diagram.  However inside the dimer phase to the left of this region ($-1\leq \beta < \beta^\ast \approx -0.2$),  we find that the disorder and Lifshitz lines merge to become a single commensurate-incommensurate (C-IC) line, i.e. $\alpha_d(\beta)=\alpha_L(\beta)\equiv \alpha_{dL}(\beta)$, as mentioned earlier in the beginning of section~\ref{sec:disorder}.  This C-IC line now marks a crossover between the commensurate dimerized phase ($S(q)$ peaked at $q=\pi$) and an incommensurate dimerized phase ($S(q)$ peaked at $q\neq\pi$), see Fig.~\ref{fig:quant-pd}. We note that this is not a true phase transition, since the spin-spin correlations are short ranged in both cases.
  Upon further increasing $\alpha$, we encounter a transition from the dimerized phase into the NNN-AKLT phase at $\alpha=\alpha_\delta$.  As we have discussed previously in section~\ref{sec:NNN}, the $\alpha_\delta$ transition is of the first order (level crossing) since the bulk gap does not close and there is no sign of divergence of the correlation length.

\section{Discussion}
\label{sec:discussion}

As stated in section~\ref{sec:NNN}, we found that the NNN-AKLT phase is distinct from the dimer phase, separated from it by a first-order phase transition $\alpha_\delta(\beta)$.
It is instructive to contrast this result with an earlier study in Ref.~\onlinecite{Pati.1997}, in which the translational symmetry of the lattice was broken by construction, by adding the term $\sum_i ((-1)^i\delta) {\bf S}_i\cdot{\bf S}_{i+1}$ to the model Eq.~(\ref{eqn:ham}) with $\beta=0$. As a result, the authors found a dimerized phase for sufficiently large $\delta$, which appeared to be smoothly connected to the NNN-AKLT phase~\cite{Kolezhuk.1996,Kolezhuk.1997-II}.  We believe that this result is a consequence of the Hamiltonian itself breaking the translational lattice symmetry, in which case the distinction between the dimer and NNN-AKLT phases becomes inessential. 

 In the present study on the other hand, the Hamiltonian (\ref{eqn:ham}) preserves translational symmetry, which becomes spontaneously broken only inside the dimer phase. Our DMRG calculations show that the NNN-AKLT phase, by contrast, preserves translational symmetry of the lattice, as manifest by the absence of the dimerization in $L\to\infty$ limit for large $\alpha$, see Fig.~\ref{fig:energy2}a and  Fig.~\ref{fig:dimer-op}a. Because of this symmetry-based distinction, and absence of the gap closing, the dimer phase and the NNN-AKLT phase must be separated by a first-order phase transition line $\alpha_\delta(\beta)$, as illustrated in Fig.~\ref{fig:schem-pd}. 
 
In addition to the DMRG analysis, one can appeal to the following argument to prove that the dimer and NNN-AKLT phases are distinct. Consider the $\alpha\to\infty$ limit (i.e. finite NNN interaction $J_2$ while $J_1 =0$), when the NNN-AKLT ground state is a good approximation to the true ground state. In this limit, the chain decouples into two independent spin-1 chains comprising odd or even sites, respectively, with an ordinary AKLT ground state in each chain as depicted schematically in Fig.~\ref{fig:chains}c. The NNN-AKLT state can thus  be thought of as two independent Haldane chains. Obviously, such a state is translationally invariant because odd and even sites are identical by construction. The dimer state, on the other hand, manifestly breaks translational symmetry as Fig.~\ref{fig:chains}b illustrates. There is also a distinction in the ground state degeneracy -- in the presence of infinitesimal $J_1$ the NNN-AKLT ground state is unique in the thermodynamic limit (with no zero-energy edge excitations). By contrast, the dimerized phase has two-fold degenerate ground state in the thermodynamic limit, corresponding to two inequivalent dimer coverings of the chain.
 
 Moreover, the NNN-AKLT state has a finite (Haldane) gap of the order of $J_2$ and as such, must be stable with respect to small perturbations, for example switching on a finite $J_1 \ll J_2$ or biquadratic interaction $|\beta J_1| \ll J_2$. Therefore, there must be a finite region in the $(\beta,\alpha)$ phase diagram where NNN-AKLT is stable. This phase can only be destroyed upon sufficiently large negative $\beta$ or large $|J_1|$. This is exactly what the DMRG phase diagram in Fig.~\ref{fig:schem-pd} shows. Of course, this argument does not tell us about the order of the phase transition in which the NNN-AKLT state is destroyed. As our DMRG calculations show, the transition from NNN-AKLT state is always first order without closing the bulk gap: either into the Haldane phase for  $\alpha<\alpha_T$, or into the dimer phase to the left of the $\alpha_\delta$ line. The only exception is the multicritical point $\Omega$ where the two lines ($\alpha_T$ and $\alpha_\delta$) meet, which is the main finding of this work.

We have not computed the string order parameter because of the limitations of the open source POWDER DMRG code~\cite{Rizzi.2008} used in this study.  Instead, we have tracked the existence or absence of the edge excitations to monitor the topological nature of the ground state.  
We have found that the edge excitations become gapped out as one crosses the critical line  $\alpha_c(\beta)$ (or first-order transition line $\alpha_T(\beta)$) from below, and  the edge wavefunction hybridizes with the bulk states. As a result, we conclude that the critical (transition) line separates the topologically nontrivial Haldane phase from the topologically trivial dimer and NNN-AKLT phases. 
It would be instructive to compute the evolution of the string order parameter across both the critical $\alpha_c(\beta)$ and transition $\alpha_T(\beta)$ lines.  It will be interesting to see whether the string order parameter goes to zero continuously across the critical line, or if it still jumps similar to the case of $\beta=0$, as earlier DMRG calculations indicate~\cite{Kolezhuk.1996,Kolezhuk.1997-II}.  

 Intriguingly, we find that the topological phase transition from Haldane into a topologically trivial phase can occur either with (for $\beta<\beta^\ast$) or without (for $\beta>\beta^\ast$) closing of the bulk gap at the transition. In the former case, the second-order transition is into the dimerized phase, whereas in the latter, the transition is of the first order into the NNN-AKLT phase without closing the bulk gap.
  While the conventional wisdom based on the bulk-edge correspondence would dictate that the bulk gap ought to close at such a transition, there is a number of examples found recently in which the bulk gap can remain finite~\cite{Ezawa.2013}. This argument provides solid footing for our findings in the region $\beta>\beta^\ast\approx -0.2$, where the DMRG calculated bulk gap remains finite across the $\alpha_T(\beta)$ line, in agreement with earlier DRMG calculations at $\beta=0$~\cite{Kolezhuk.1996,Kolezhuk.1997-II} and field-theoretical results~\cite{Allen.1995}.

We have established the existence of both a critical line $\alpha_c(\beta)$ and transition line $\alpha_T(\beta)$ in the phase diagram of Fig.~\ref{fig:schem-pd}. Crucially, crossing a line of the phase boundary does not require fine-tuning of both $\alpha$ and $\beta$ parameters, as opposed to a single point in the phase diagram.
 Experimentally,  both $\alpha$ and $\beta$ interactions are generically present in the system, and our results thus raise an exciting prospect of being able to observe the critical/transition line experimentally. Therefore, we expect that a signature of the transition may be accessible in an experiment, for not too large values of $\alpha$ and $\beta$. This is in contrast to, say, the TB point or the ULS point, which would require fine-tuning and a large value of the biquadratic interaction $|\beta|=1$, which is likely  unrealistic. 

This work has focused on the parameter range $\beta<1$ and $\alpha>0$.  It will be very interesting to consider the effect of frustration on the ULS point and the gapless~\cite{Lauchli.2006} antiferro-quadrapolar phase for $\beta >1$.  It is a natural question to ask whether this phase will still exist for a finite $\alpha$, and if so, whether there is a direct transition between this phase and the NNN-AKLT phase on the right-hand side of the phase diagram of Figure~\ref{fig:schem-pd}.   In addition, the ULS point $(\beta=1,\alpha=0)$ is known to be described by the SU$(3)_{k=1}$ CFT with gapless modes at $q = 0, \pm 2\pi/3$ (that show up clearly in $S(q)$, see Ref.~\onlinecite{Bursill.1995}).  It will be very interesting to consider perturbations (as a result of a finite $\alpha$) to the ULS point, similar to the field theoretical approach in section~\ref{sec:field-theory}.  Also, in this case the model has already passed through a disorder and Lifshitz point due to the large positive $\beta$, and we therefore expect  the peaks of the spin structure factor $S(q)$ to shift continuously with increasing $\alpha$, from $\pm 2\pi/3$ to $\pm \pi/2$.

\section{Conclusions}
\label{sec:conclusions}

By combining field theoretical arguments, DMRG calculations, and an analytic variational wavefunction ansatz with $M=4$ kept states, we have mapped out the phase diagram of the frustrated antiferromagnetic spin-1 chain in terms of the biquadratic spin interaction ($\beta$) and next-nearest neighbor  exchange ($\alpha>0$). Our results smoothly connect previous studies along the isolated lines ($\alpha=0$ or $\beta=0$) in the phase space of parameters, and provide a unified physical picture of the spin-1 chain model in the entire $\alpha-\beta$ plane.
 We identify three main phases: the Haldane  phase, the next-nearest-neighbor AKLT phase, and the dimerized phase. We found, for the first time, that the dimerized and Haldane phases are separated by a line of second order phase transitions that originates at the well studied Takhtajan--Babudjian point ($\beta\!=\!-1$, $\alpha\!=\!0$), and terminates at a previously unidentified multicritical point $\Omega\!=\!(\beta^\ast$, $\alpha^\ast)$, with approximate coordinates $-0.2<\beta^\ast<-0.15$ and $0.47<\alpha^\ast < 0.53$.
 Based on field-theoretical analysis, we propose that the conformal field theory describing the low-energy excitations at the $\Omega$ point is distinct from previously known gapless points in the phase diagram of the spin-1 chain, and is characterized by the SU$(2)_4$  Wess-Zumino-Witten theory with central charge $c=2$. This conclusion is corroborated by the DMRG calculated central charge, deduced from the finite-size scaling of the entanglement entropy.
 
 To the right of the multicritical point (for $\beta>\beta^\ast$), the critical line becomes a line of first order phase transitions, corroborating earlier DMRG calculations~\cite{Kolezhuk.1996,Kolezhuk.1997-II} at $\beta=0$. This first order transition line separates the NNN-AKLT and Haldane phases. Since the Haldane phase can be understood as a symmetry-protected topological phase~\cite{Gu.2009, Pollmann.2012}, this is an example of a topological phase transition that occurs without closing of the bulk gap. We also provide numerical evidence that the dimer and NNN-AKLT are two distinct (topologically trivial) phases, separated from each other by a line of first-order phase transitions.
 These findings are corroborated by DMRG calculations of the bulk and edge gaps, spin-spin correlation length,  ground state energy, and the dimer order parameter.  In addition, we have used an analytical matrix product state anzats for a variational wavefunction, which allowed us to determine semi-quantitatively various phases and transitions between them and provided a useful intuitive guide to the DMRG calculations. 

Prior to this work, quantum transitions between different phases in the spin-1 chain have only been seen theoretically at isolated fine-tuned points, making an experimental realization very challenging.  Here, we have established the existence of  several \emph{lines} of phase transitions, which do not require careful tuning of the parameters and therefore  should be more readily accessible in experiments on quasi-one-dimensional materials. Another possible realization may be found in ultracold atoms, where there are proposal to artificially engineer spin chains using spinor atoms in an optical lattice.\cite{Yip.2003, Imambekov.2003, Ripoli.2004} 

 In addition to the aforementioned three distinct phases (Haldane, NNN-AKLT and dimer), we also used DMRG to identify regions of the phase diagram characterized by various short-range orders in the spin-spin correlation function. 
Extending earlier DMRG work by other authors\cite{Schollwock.1996,Kolezhuk.1996,Kolezhuk.1997-II},  we show the existence of two incommensurate crossovers inside the Haldane phase: the Lifshitz transition $\alpha_L$ and the so-called disorder transition of the first kind $\alpha_d$, marking incommensurate correlations in momentum and real space, respectively. Whereas earlier, these two transitions have been only characterized at isolated points ($\alpha=0$ or $\beta=0$), here we show that they stretch across the entire $(\alpha,\beta)$ phase diagram.
 Inside the dimer phase these two lines  merge into a single incommensurate-to-commensurate transition line. This behavior is similar to that seen in classical frustrated two-dimensional spin models.   Intriguingly, we find that the point of this merging coincides with the multicritical point $\Omega$, at least within the precision of our numerical calculations. 
The existence of this multicritical point in the phase diagram, where the Haldane, dimer and NNN-AKLT phases meet, is conceptually perhaps the most important finding of this work.

\section{Methods}
\label{sec:methods}
\subsection{DMRG}
For the DMRG calculations presented here we are using the open source POWDER DRMG code~\cite{Rizzi.2008}.  We extrapolate our results in system sizes for $L=40,50,60,70,80,90,$ and $100$ and have also considered various different numbers of kept states ranging from $M=80,120,160,$ and $200$ .  For a fixed number of kept states we determine the ground state energy, the gap and the dimer order parameter in the thermodynamic limit by fitting to quadratic polynomials
\begin{eqnarray}
f(L,M) &=& f({\infty},M) + a_f(M)/L +b_f(M)/L^2,
\label{eqn:thermo-limit}
\end{eqnarray}
$f$ is quantity being extracted to the the thermodynamic limit with the extrapolated value $f(L={\infty},M)$.  In addition we have also studied the convergence in $M$ for the ground state energy and the gap as a function of $\alpha$ for $\beta=-0.2$ and $0.2$.  We extrapolate our results in $M$ from
\begin{eqnarray}
f(\infty,M) &=& f^{\infty} + a_f^*/M .
\label{eqn:thermo-limit2}
\end{eqnarray}
 after the extrapolation in $L$ (similar to Refs.~\onlinecite{Rosengren-I,Rosengren-II,Capone}), once each quantity develops a linear dependence on $1/M$.  For $\beta>0$ we present correlation functions for chain lengths $L=100$ and for $\beta<0$ we used $L=80$ with $M=200$ kept state for both, such that the truncation error is at most $10^{-9}$ (when we are away from the critical and transition lines) and perform 5 finite size DMRG sweeps.  In the vicinity of the critical and transition lines the truncation error can be as large as $10^{-7}$.  
  
\begin{figure}[t!]
\includegraphics[width=3.in]{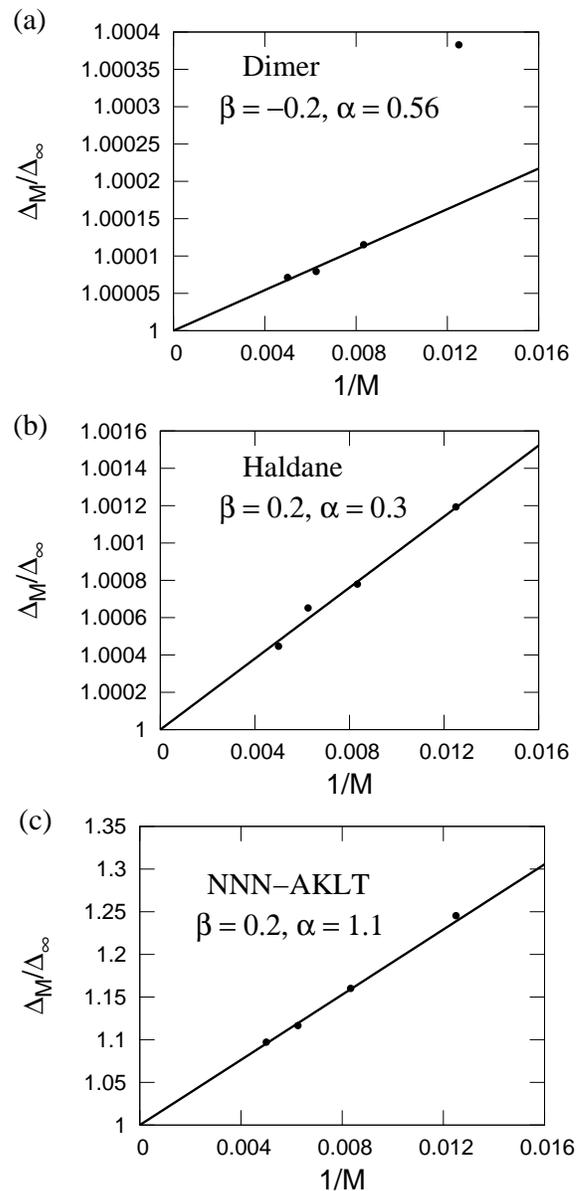}
\caption{(Color online)  The dependence of the gap $\Delta$, on the number of kept states $M$ in each relevant phase (a) Dimer, (b) Haldane, and (c) NNN-AKLT. }
\label{fig:M-dep-gap}
\end{figure}

\begin{figure}[t!]
\includegraphics[width=3.in]{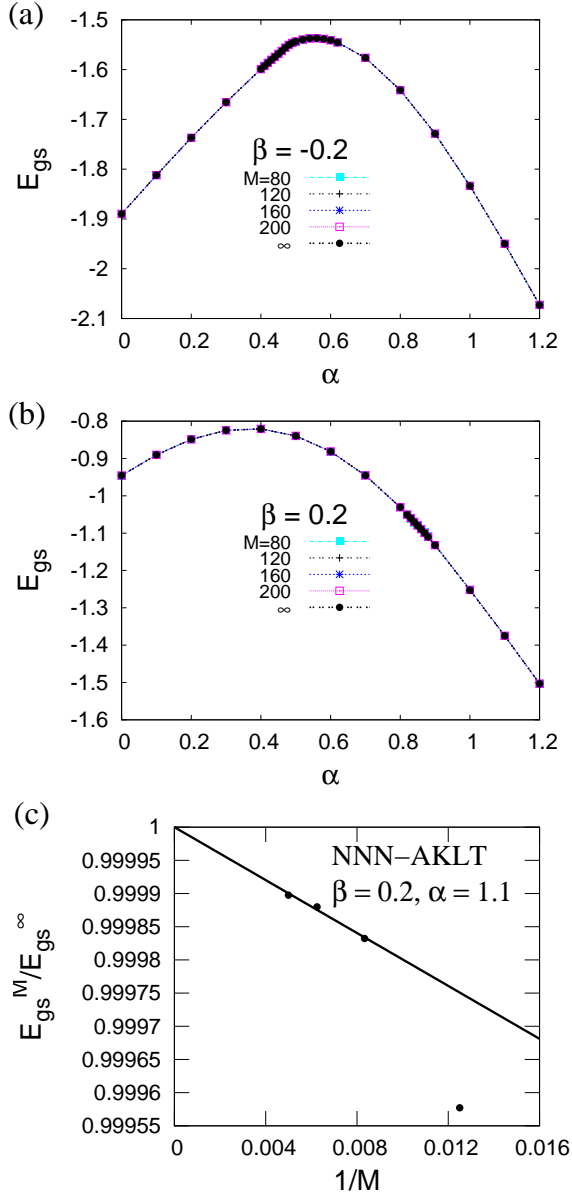}
\caption{(Color online)  The ground state energy $E_{gs}$, for various different number of kept states $M$ for (a) $\beta=-0.2$ and (b) $\beta=0.2$ as a function of $\alpha$.  In contrast to the gap (Fig.~\ref{fig:M-dep-gap}), the $M$ dependence of $E_{gs}$ is much weaker. (c) The most significant $M$ dependence of the ground state energy is observed in the NNN-AKLT phase. }
\label{fig:M-dep-Egs}
\end{figure}

We find the dimer and Haldane phases are reasonably well converged in $M$ even at $M=80$, where going to larger values of $M$ results in a small shift in the numerical value of the gap (see Figs. \ref{fig:xi-gap2} and \ref{fig:xi-gap1} for various values of $M$ as a function of $\alpha$ and Figs. \ref{fig:M-dep-gap}a  and \ref{fig:M-dep-gap}b  for the explicit $M$ dependence).  This is quite natural since the ground state wavefunction in these phases has a relatively simple valence bond like structure and as a result each phase is minimally entangled.  
Interestingly, we find a very weak $M$ dependence even in the vicinity of transition between the Haldane and Dimer phases, see Fig. \ref{fig:xi-gap2}a.  By contrast, in the NNN-AKLT phase the gap has a significant dependence on $M$ (see Figs. \ref{fig:xi-gap2}a and \ref{fig:xi-gap1}a for various values of $M$ as a function of $\alpha$ and Fig.~\ref{fig:M-dep-gap}c for the explicit $M$ dependence) and our results even at $M=200$ are not well converged, making an extrapolation in $1/M$ necessary.  Such a dependence on $M$ was discussed in the study of the frustrated spin $S=1$ chain $(\beta=0$) Ref.~\onlinecite{Kolezhuk.1997-II}, which is due to the ground state wavefunction in the  NNN-AKLT  phase being, in a sense, a direct product of two AKLT wavefunctions and therefore requiring a significant number of more kept states.  As a result we find a significant $M$ dependence of DMRG results in the vicinity of the transition into the NNN-AKLT phase, see Fig. \ref{fig:xi-gap1}a.  In each phase we find the ground state energy calculations are more well converged in $M$ then the gap, compare Figs. \ref{fig:xi-gap2}a and \ref{fig:xi-gap1}a with Figs. \ref{fig:M-dep-Egs}a,b and Figs. \ref{fig:M-dep-gap}c and \ref{fig:M-dep-Egs}c at $\beta=0.2$ and $\alpha=1.1$ in the NNN-AKLT phase.  

\subsection{Conformal Field Theory}\label{sec:CFT}

In this section, we provide a more rigorous field theoretical argument regarding the statement made in Section~\ref{sec:CFT1} about the conformal field theory SU(2)$_4$ which, we conjecture describes the critical end-point $\Omega$ in the ($\alpha,\beta$) phase diagram (see Figs.~\ref{fig:schem-pd}, \ref{fig:flow}).

We start from the representation of the spin-1 Hamiltonian in terms of two coupled chains as in Eq.~(\ref{eqn:chains}). Even though in this work we are  interested in an antiferromagnetic spin chain, it is instructive to consider the case of a ferromagnetic interchain coupling $J_\perp$ (recall that $J_\perp$ in Eq.~(\ref{eqn:chains}) corresponds to $J_1$ in the original model Eq.~\ref{eqn:ham}). Let us consider the dimerized phase above the solid line in the phase diagram Fig.~\ref{fig:flow}. Then, for sufficiently large ferromagnetic $J_\perp$, the alternating rungs will form spin-triplet dimers as in Fig.~\ref{fig:chains}b. The model is then equivalent to a spin $S=2$ chain, in the limit when the coupling $J_2$ is not too large. The question now is: what conformal field theory describes a critical point in this $S=2$ model?

Because of the SU(2) spin symmetry, the sought theory is most likely the WZW SU(2)$_k$ at level $k$. It is well established that the Kac-Moody currents $J_L^a, J_R^a$ of such a theory can be expressed as bilinears in terms of free (albeit nonlocal) fermionic  degrees of freedom~\cite{CFT.book, Tsvelik.book}. In particular, $J_R^a$ is expressed through the right (R) movers: $J_R^a = \sum_n :\!\psi^\dagger_{R,\alpha,n} \tau_{\alpha\beta}^a \psi_{R,\beta,n}\!\!:$, with a similar expression for $J_L^a$ in terms of left movers, where $\tau^a$ are the generators of the SU(2) group (Pauli matrices). 
Consider now 2 neighboring effective spins $S=2$: we need $k=4$ ``flavours'' of fermions to describe all the degrees of freedom. Such a free fermionic theory has U($k$)$\cross$ SU($N$) symmetry. Using a group identity U$(k) \cross$SU($N$) = U(1)$\cross$SU$(N)\cross$SU($k$), one can represent the fermionic operators in terms of the product of an SU$(2)_k$ spin WZW field, an SU$(k)_2$ flavour WZW field, and a free boson corresponding to the charge U(1) field~\cite{Affleck.1986, Affleck.1991}. Since we are dealing with a charge insulator, the U(1) field is gapped. The flavour field corresponds to the ``valley'' SU($k$) symmetry and is of no consequence (in fact, it can be gapped out by introducing perturbations to the model~\cite{Tsvelik.book}). The only remaining gapless field describes the low-energy spin-2 degrees of freedom in terms of the SU(2)$_{k=4}$ WZW theory. This is an example of a more general proposal by Affleck~\cite{Affleck.PRL.1986} that certain multicritical points of the spin-$S$ Heisenberg model are described by SU(2)$_k$ WZW theory at level $k=2S$.

As noted in Section~\ref{sec:CFT1}, the SU(2)$_4$ theory has one relevant field with scaling dimension $\Delta_1=2/3$, expressed in terms of the bilinear of primary operators, as well as two marginal  fields -- one relevant and one irrelevant. It is instructive to analyze what those fields correspond to in terms of the original spin model. For this, let us consider in more detail the two coupled spin chains in Eq.~(\ref{eqn:chains}). 
Each  spin-1 chain in (\ref{eqn:chains}) can be described by the WZW theory with perturbation as in Eq.~(\ref{eqn:lagrange}):
\beq
\mathcal{L} = \mathcal{L}_1^{\text{WZW}} + \mathcal{L}_2^{\text{WZW}}  + \mathcal{L}_\perp+ \{\text{perturbations}\}.
\eeq
Following Allen and S\'en\'echal~\cite{Allen.2000}, the coupling between the two chains can be described in the language of the aforementioned primary fields $\hat{g}^a$ and the conformal currents $J^a, \bar{J}^a$ of each chain:
\bea
\mathcal{L}_\perp &=& \lambda_2 (J_1^a J_2^a + \bar{J}_1^a \bar{J}_2^a ) + \lambda_3 (J_1^a \bar{J}_2^a +  \bar{J}_1^a J_2^a) \nonumber \\
 &+& \rho \left[ g_1^a (\pd_x g_2^a) - (\pd_x g_1^a) g_2^a\right].
\label{eqn:Hperp}
\eea
The last term, referred to as the \emph{twist term} is strongly relevant and is highly nontrivial to analyze due to its non-vanishing conformal spin. The bosonization treatment by Allen and S\'en\'echal~\cite{Allen.2000} showed that it is responsible for the onset of incommensurability in the spin-spin correlations as $J_\perp$ increases. It is this relevant term that leads to the dimer or NNN-AKLT phase and opens up the spectral gap, and we can therefore relate it with the corresponding \emph{relevant} bilinear of the primary operators in the effective SU(2)$_4$ theory.

The first two terms in Eq.~(\ref{eqn:Hperp}) have a scaling dimension $\Delta=2$ and are therefore marginal. Their effect depends on the sign of the coupling constants: both $\lambda_2$ and $\lambda_3$ are proportional to $J_\perp\equiv J_1$ and become \emph{relevant} for an antiferromagnetic $J_\perp$. Indeed, in this case  the lowest-lying excitations are the rung triplets on the zigzag chain in Fig.~\ref{fig:chains} and they cost an energy $\sim J_\perp$ for large positive interchain coupling~\cite{Allen.2000}. 
In the language of the effectibe $S=2$ model, this is the \emph{marginally relevant} field of the SU(2)$_4$ model.

Finally, there remains a \emph{marginal} current bilinear  in the SU(2)$_4$ model. It corresponds to the marginally irrelevant current bilinear $(\bar{J}_nJ_n)$ in each of the individual chains ($n=1,2$), which only becomes marginally relevant if the intra-chain interaction $J_2$ is ferromagnetic. In this work, we only consider antiferromagnetic $J_2$, so this term remains \emph{marginally irrelevant}.

To conclude, we have shown that the spin-1 chain exhibits emergent $S=2$ excitations and provided those are gapless, they are described by the critical SU(2)$_4$ conformal field theory. We have identified the physical meaning of various relevant and marginal operators of this theory by establishing their correspondence with the fields of the two coupled spin-chain model in Eq.~(\ref{eqn:chains}).

\subsection{Variational MPS wavefunction}\label{sec:MPS}

\begin{figure}[h!]
\includegraphics[scale=.8]{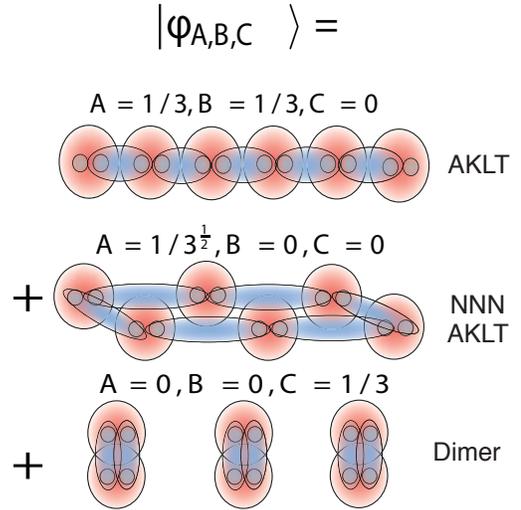}
\caption{The variational Matrix Product State (MPS) wavefunction is constructed using three states, AKLT, NNN-AKLT, and dimer, which can be represented individually as MPSs, see Eq.~(\ref{eq:MPS-var-wf}). Parameters $A,B,C$ can be chosen to interpolate between the three states, which can each be recovered by a special choice of parameters shown in the figure above.}\label{fig:MPS-wf-schem}
\end{figure}

Matrix Product States (MPS) have emerged as a powerful tool for analytic studies of correlated quantum systems. The representation of the exact ground state of the one-dimensional AKLT state using MPS~\cite{Affleck.1987} led to in-depth studies of these states \cite{Fannes.1989}, which pertinent to the present case, were subsequently used for analytical calculations for spin$-1$ chains, e.g. in Refs.~\cite{Klumper.1993,Kolezhuk.1996,Kolezhuk.1997-II,Schollwock.1996,Kolezhuk.1997-I,Kolezhuk.1998,Kolezhuk.1999}. In this tradition, we construct a variational MPS wavefunction first introduced in Refs.~\onlinecite{Kolezhuk.1996, Kolezhuk.1997-II}, to study the ground state and low-lying excitations of Hamiltonian (\ref{eqn:ham}), depicted in Fig.~\ref{fig:MPS-wf-schem}.
We first consider the MPS representation of the AKLT wavefunction, $|\psi\rangle = {\rm Tr}\left[g_1 \cdot g_2\cdot g_3\cdot \dots \cdot g_M\right]$ with
\beq
 g_{\text{AKLT}}= \frac{1}{\sqrt{3}} \left( \begin{array}{cc}
|0\rangle & -\sqrt{2}|+\rangle  \\
\sqrt{2}|-\rangle & -|0\rangle \end{array} \right) \label{eqn:aklt_mps}
\eeq
where the matrices $g_i$ at sites $i=1,2,...M$ are identical due to translational invariance, and each is constructed in the $S=1$ local spin basis at site $i$; the construction can be understood by first realizing that the AKLT Hamiltonian can be rewritten as a sum of projectors onto $S=2$ in the space of spins at adjacent sites $i,i+1$. This implies that the ground state must have total spin $S_{i,i+1} \neq 2$ at sites $i,i+1$. Such a state can be conceptualized by introducing a set of auxiliary  $S=1/2$ spins at each site of the chain (see Fig.~\ref{fig:schem-states}), and preparing each adjacent pair of $S=1/2$ spins in a singlet. It is simple to check that the sequence of matrices $g$ in Eq.(\ref{eqn:aklt_mps}) encapsulates this structure. 

\subsubsection{Variational approach to the ground state}

The AKLT MPS is a good approximation to the true ground state of $H$ in the vicinity of the AKLT point $\beta=1/3,\alpha=0$ (see \cite{Affleck.1987,Kolezhuk.1996}). To go beyond this regime we use the physical insight that next-nearest neighbor interactions dominate over nearest-neighbor ones for large positive $\alpha$ and that the system dimerizes for large negative $\beta $. Thus, we introduce the ``next-nearest-neighbor" AKLT and the ``dimerized" MPS which contain these two essential effects, respectively. In the spirit of Refs.~\onlinecite{Kolezhuk.1996, Kolezhuk.1997-II} we construct a variational MPS which interpolates between the AKLT MPS and the two additional MPS described above (see Fig.~\ref{fig:MPS-wf-schem}): 
\bea
\label{eq:MPS-var-wf}
|\phi_{A,B,C}\rangle &=& {\rm Tr}\left[\Gamma_{1,2} \Gamma_{3,4} \Gamma_{5,6}...\Gamma_{M-1,M}\right], \\
\Gamma_{1,2} &=& \sum_{i,j} |t_{1i},t_{2j}\rangle \bigg[ A\delta_{i,j} \mathbb{1}_4 + i B \varepsilon_{ijk}(\sigma_k\otimes\mathbb{1}_2)\nonumber\\
&&+i C (\sigma_i \otimes \sigma_{j})\bigg]
\eea
where $A,B,C$ determine the relative weights of the three candidate MPS states, $\sigma_{i}$ are usual 2$\times$2  Pauli spin matrices, $\varepsilon_{ijk}$ is the Levi-Civita symbol, and the matrices $\Gamma_{i,i+1}$ are identical due to translational invariance. Note that it is more convenient to construct the variational state in the space of 2 adjacent spins, concretely $|t_{1i} t_{2j}\rangle$ is the decoupled basis of $S=1$ spins at sites 1,2, with $|t_i\rangle$ expressed in the Cartesian basis $i = (x,y,z)$, i.e. 
\bea
S_x|t_x\rangle =0, \  \ S_y|t_y\rangle = 0, \ \  S_z|t_z\rangle = 0.
\eea

Choosing parameters $A=B=1/3, C=0$ yields the AKLT MPS, while the completely dimerized state is given by $A=1/\sqrt{3},B=C=0$, and the NNN-AKLT state corresponds to $A=B=0,C=1/3$. The crucial point is that optimal $A,B,C$ can be determined by minimizing the energy $\langle \phi|H|\phi\rangle$. 

It is most convenient to compute expectation values of local observables in the MPS using the transfer matrix technique of Refs.~\cite{Klumper.1993,Totsuka.1995,Kolezhuk.1996,Kolezhuk.1997-II}. For example the total energy can be decomposed into a sum of local operators $H = \sum_{i} h^{nn}_{i,i+1} + h^{\rm nnn}_{i,i+2}$ which can then individually be evaluated in state $|\phi\rangle$. The essential steps involve computing and diagonalizing the transfer matrix $G=\Gamma^\dagger \otimes \Gamma$, a 16$\times$16 matrix; all operator expectation values will involve traces over chains of the matrix $G$ sandwiched between operators, e.g.
\bea
& &\langle O^{(1)}(i)O^{(2)}(j)\rangle=
\nonumber
\\
& &{\rm Tr}[G_1.... O^{(1)}(i)G_{i+1}....O^{(2)}(j)....G_M],
\eea
which in the thermodynamic limit will be dominated by the largest eigenvalue of $G$ leading to
\bea
& &\lim_{M\to\infty} \langle O_{i}O_{j}\rangle  =
\nonumber
\\
& & \lambda_{\rm max}^{M} \sum_m \lambda_{m}^{j-i-1}(U^\dagger O^{(1)}U)_{1,m}(U^\dagger O^{(2)}U)_{m,1},
\label{eq:mps_obs_tdl}
\eea 
where $\lambda_{\rm max}$ is the largest eigenvalue of $G$. $U$ diagonalizes $G$, i.e. $U^\dagger G U = \lambda_i \delta_{ij}$ with the convention $\lambda_1 = \lambda_{\rm max}$. We find it convenient to normalize the maximal eigenvalue of $G$ to unity which leads to a constraint~\cite{Kolezhuk.1996,Kolezhuk.1997-II} on $A,B,C$:
\bea
3A^2 + 6B^2 + 9C^2 =1.
\label{norm}
\eea
Observables including two-point correlation functions can be calculated either directly using Eq.~(\ref{eq:mps_obs_tdl}) or by straightforward generalizations. 
Thus within the variational approach we can calculate the ground state energy as well as correlation functions. 

\subsubsection{Variational approach to the low energy excitations}

It is also possible to approximate the low lying excitations above the ground state within the MPS framework \cite{Solyom,Kolezhuk.1997-II}. The idea shadows the single mode approximation used to obtain dispersions for spin chains \cite{assabook}: excitations are constructed by adding traveling ``defects'' or solitons \cite{Solyom,Kolezhuk.1997-II} to the ground state,
\bea
\label{eq:MPS_SMA}
|E(k)\rangle = \sum_{n}e^{i n k} {\rm Tr}\left[g_1\otimes g_2\otimes... c_n\otimes... g_M\right],
\eea
where $c_n$ is a defect at site $n$, which e.g. in the AKLT case is given by
\bea
c_n = \sigma^{\pm,z} g_{\rm AKLT}.
\eea
Note that there are three possible excitation modes which can be generated by the appropriate spin defect; here it corresponds to choosing one of the three Pauli sigma matrices in the spherical basis $(z,+,-)$ \cite{Kolezhuk.1997-II}. Due to the rotational symmetry of our model, the three dispersion modes are degenerate.

\begin{figure}[h!]
\includegraphics[width=3.in]{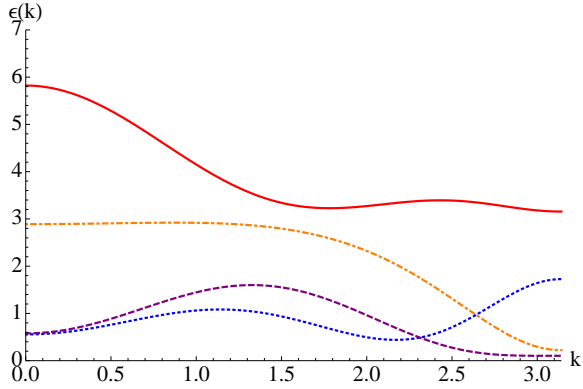}
\caption{(Color online) Approximate dispersion curves $\varepsilon(k)$ calculated for representative choices of couplings $\alpha, \beta$ within the single mode or ``crackion'' approach. The dotted line (blue) is evaluated for $\alpha = 0, \beta = 0$, is shown for reference and captures the physics deep within the AKLT phase; the dashed-dotted (orange) line corresponds to $\alpha > 0$ and approximates the excitations in the NNN-AKLT phase; the dashed (purple) line corresponds to $\beta < 0$, the dimer phase, and the solid (red) line corresponds to $\beta > 0$. We observe a shift in the minimum of the dispersion curve away from the AKLT phase, in which the minimum occurs at $k = \pi$ as we tune $\alpha, \beta$, see also Fig.~\ref{fig:chiq}.}
\label{fig:crack-disp}
\end{figure}

\begin{figure}[h!]
\includegraphics[scale=0.95]{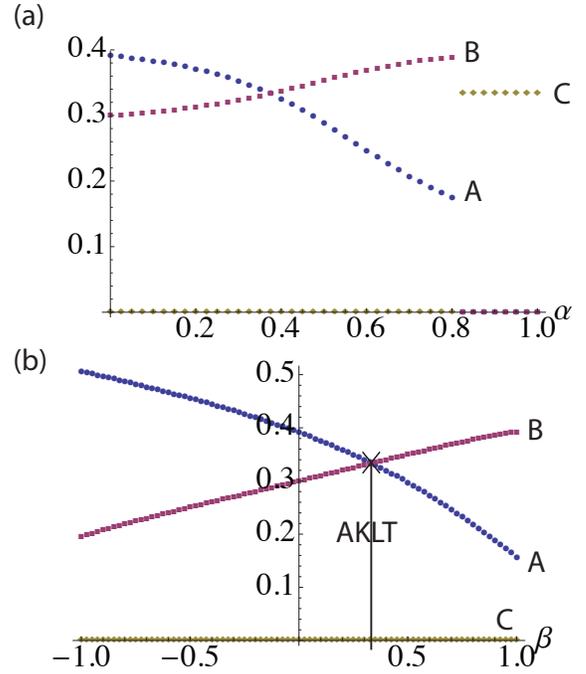}
\caption{(Color online) Values of $A,B,C$ (respectively blue circles, red squares, yellow diamonds) shown as next-nearest neighbor coupling $\alpha$ and biquadratic coupling $\beta$ are tuned across the pure Heisenberg coupling point $\alpha = 0, \beta = 0$ deep inside the Haldane phase. (a) For a fixed $\beta = 0$ we show $A,B,C$ as a function of $\alpha$; (b) For a fixed $\alpha = 0$ we show $A,B,C$ as a function of $\beta$ -- the point $\beta = 1/3$ corresponds to the AKLT point for which the AKLT MPS state is the exact ground state (corresponding to $A = B = 1/3, C = 0$. Note that, within our approximation, the AKLT state also coincidentally describes the point $\alpha = 0.374, \beta = 0$ shown in (a). Moreover, we observe that the parameters $A,B,C$ are ``close'' to the AKLT values for a large portion of the phase diagram (with the exception of regions deep in the NNN-AKLT phase $\alpha \gtrsim 0.8$ ).}
\label{fig:ABC-typ}
\end{figure}

To gain intuition into the structure of the low-energy excitations in the three phases depicted in Fig.~\ref{fig:schem-pd}, within the approximation of Eq.~(\ref{eq:MPS_SMA}) we calculated the dispersion $\varepsilon(k)$ of the AKLT MPS state which we obtained in closed form (see also Refs.~\onlinecite{Kolezhuk.1996,Kolezhuk.1997-II}):
\bea
\varepsilon(k) &=& \frac{14}{9} + \frac{26}{27}\alpha + \frac{160\alpha - 18}{27} \cos(k) - \frac{14}{9}\alpha \cos(2k) \nonumber\\
&&+\left(2 - \frac{26}{3}\alpha\right)\frac{3 + 5 \cos(k)}{5 + 3 \cos(k)} + \frac{2}{9}\beta (13 + 9 \cos(2 k)).\nonumber\\
\eea
By choosing values of $\alpha, \beta$ lying in the relevant regions of the phase diagram in Fig.~\ref{fig:schem-pd}, we obtain dispersion curves which we argue characterize the low energy excitations of the AKLT, NNN-AKLT, and dimer phases. Thus we present the spin dispersion within this approximation for representative values of $\alpha, \beta$ in Fig.~\ref{fig:crack-disp}. The justification for using the AKLT MPS ground state as the starting point to explore dispersions in neighboring phases is provided by the fact that even within our variational approach, the {\em ground states} in a region of the NNN-AKLT and dimer phases are well approximated by the AKLT state (see Fig.~\ref{fig:ABC-typ}).

\section{Acknowledgements}
We would like to thank Marton Kormos, Matthew Foster, Ian Affleck, Steven White,   T. Senthil,  Karen Hallberg, Oleg Starykh, Stephan Deppenbrock, Vadim~Oganesyan, Collin Broholm, Maissam Barkeshli, Qimiao Si, Wenxin Ding, and Yang-Zhi Chou for useful discussions.  We would especially like to thank Matteo Rizzi for his detailed help in using the POWDER DMRG code and to I. Affleck for bringing Ref.~\onlinecite{Affleck.PRL.1986} to our attention. We are indebted to Oleg Tchernyshyov and  Masaki Oshikawa for discussions concerning the distinction between the dimer and NNN-AKLT phases. The calculations have been performed on Rice University Data Analysis and Visualization Cyberinfrastructure funded by the NSF under grant OCI-0959097.  JHP was supported in part by the East-DeMarco fellowship; AHN was supported by the Welch Foundation grant C-1818 and Cottrell Scholar Award from Research Corporation for Science Advancement. AHN acknowledges the hospitality of the Aspen Center for Physics supported by the NSF grant no. 1066293, where part of this work was performed. 
\clearpage


\end{document}